\documentclass[twocolumn, amsmath,amssymb,pra]{revtex4}
\pdfoutput=1

\usepackage{graphicx}
\usepackage{dcolumn}
\usepackage{bm}
\usepackage{bbm}
\usepackage{color}
\usepackage{url}
\usepackage[normalem]{ulem}

\newcommand{\be}{\begin{equation}}
\newcommand{\ee}{\end{equation}}
\newcommand{\bd}{\begin{displaymath}}
\newcommand{\ed}{\end{displaymath}}
\newcommand{\BE}{\begin{eqnarray}}
\newcommand{\EE}{\end{eqnarray}}

\newcommand{\bra}{\left\langle}
\newcommand{\ket}{\right\rangle}

\newcommand{\bs}{\ensuremath{\mathbf{s}}}
\newcommand{\bu}{\ensuremath{\mathbf{u}}}
\newcommand{\bv}{\ensuremath{\mathbf{v}}}

\newcommand{\mcD}{\mathcal{D}}
\newcommand{\mcM}{\mathcal{M}}

\begin{document}
\preprint{}
\title{Estimation of effective temperatures in quantum annealers for sampling applications: A case study with possible applications in deep learning}

\author{Marcello Benedetti}
\affiliation{Quantum Artificial Intelligence Lab., NASA Ames Research Center, Moffett Field, CA 94035, USA}
\affiliation{SGT Inc., 7701 Greenbelt Rd., Suite 400, Greenbelt, MD 20770, USA}
\affiliation{Department of Computer Science, University College London, WC1E 6BT London, United Kingdom}
\author{John Realpe-G\'omez}
\affiliation{Quantum Artificial Intelligence Lab., NASA Ames Research Center, Moffett Field, CA 94035, USA}
\affiliation{SGT Inc., 7701 Greenbelt Rd., Suite 400, Greenbelt, MD 20770, USA}
\affiliation{Instituto de Matem\'aticas Aplicadas, Universidad de Cartagena, Bol\'ivar 130001, Colombia}
\author{Rupak Biswas}
\affiliation{Exploration Technology Directorate, NASA Ames Research Center, Moffett Field, CA 94035, USA}
\author{Alejandro Perdomo-Ortiz}
\email{alejandro.perdomoortiz@nasa.gov}
\affiliation{Quantum Artificial Intelligence Lab., NASA Ames Research Center, Moffett Field, CA 94035, USA}
\affiliation{University of California Santa Cruz @ NASA Ames Research Center, Moffett Field, CA 94035,
USA}
\begin{abstract}
An increase in the efficiency of sampling from Boltzmann distributions would have a significant impact on deep learning and other machine-learning applications. Recently, quantum annealers have been proposed as a potential candidate to speed up this task, but several limitations still bar these state-of-the-art technologies from being used effectively. One of the main limitations is that, while the device may indeed sample from a Boltzmann-like distribution, quantum dynamical arguments suggest it will do so with an {\it instance-dependent} effective temperature, different from its physical temperature. Unless this unknown temperature can be unveiled, it might not be possible to effectively use a quantum annealer for Boltzmann sampling. In this work, we propose a strategy to overcome this challenge with a simple effective-temperature estimation algorithm. We provide a systematic study assessing the impact of the effective temperatures in the learning of a special class of a restricted Boltzmann machine embedded on quantum hardware, which can serve as a building block for deep-learning architectures. We also provide a comparison to $k$-step contrastive divergence (CD-$k$)  with $k$ up to 100. Although assuming a suitable fixed effective temperature also allows us to outperform one step contrastive divergence (CD-1), only when using an instance-dependent effective temperature do we find a performance close to that of CD-100 for the case studied here.
\end{abstract}

\maketitle

\section{Introduction}

The use of quantum computing technologies for sampling and machine learning applications has attracted increasing interest from the research community in recent years \cite{neven2008training, neven2009nips, neven2009training, bian2010ising, Denil-2011,denchev2012robust, Lloyd-arXiv-2013,Pudenz-QIP-2013, Dumolin-2014,Lloyd-NatPhys-2014,Rebentrost-PRL-2014, Wiebe-arXiv-2015, Aaronson-2015,Adachi-arXiv-2015, chancellor2016maximum,Amin-arXiv-2016}. Although the main focus of the quantum annealing computational paradigm~\cite{finnila1994quantum,kadowaki_quantum_1998,Farhi2001} has been on solving discrete optimization problems in a wide variety of application domains~\cite{Gaitan2012,PerdomoOrtiz2012_LPF, Bian2014, OGorman-EPJST-2015,RieffelQIP2015,PerdomoOrtiz_EPJST2015,perdomo2015performance, Venturelli-JobShop-arXiv-2015}, it has been also introduced as a potential candidate to speed up computations in sampling applications. Indeed, it is an important open research question whether or not quantum annealers can sample from Boltzmann distributions more efficiently than traditional techniques~\cite{bian2010ising, Denil-2011,Dumolin-2014}.

There are challenges that need to be overcome before uncovering the potential of quantum annealing hardware  for sampling problems. One of the main difficulties is that the device does not necessarily sample from the Boltzmann distribution associated with the physical temperature and the user-specified control parameters of the device. Instead, there might be instance-dependent corrections leading, in principle, to instance-dependent effective temperature~\cite{bian2010ising,Amin-arXiv-2015,PerdomoOrtiz_SciRep2016}. Bian {\it et al.} \cite{bian2010ising} used the maximum-likelihood method to estimate such an instance-dependent temperature and introduced additional shifts in the control parameters of the quantum device; this was done for several realizations of small eight-qubit instances on an early generation of quantum annealers produced by D-Wave Systems. The authors showed that, with these additional estimated shifts in place, the empirical probability distribution obtained from the D-Wave appears to correlate very well with the corresponding Boltzmann distribution. Further experimental evidence of this effective temperature can be found in Ref.~\cite{PerdomoOrtiz_SciRep2016} where its proper estimation is needed to determine residual bias in the programmable parameters of the device.

Recent works have explored the use of quantum annealing hardware for the learning of Boltzmann machines and deep neural networks \cite{Adachi-arXiv-2015, Dumolin-2014, Denil-2011, bian2010ising, Dorband}. Learning of a Boltzmann machine or a deep neural network is in general intractable due to long equilibration times of sampling techniques like Markov chain Monte Carlo (MCMC)~\cite{Sinclair-InfComp-1989,Frigessi-Biometrika-1997,Long-2010}. One of the strategies that have made possible the recent spectacular success~\cite{LeCun-Nature-2015} of these techniques is to deal with less general architectures that allow for substantial algorithmic speedups. Restricted Boltzmann machines (RBMs)~\cite{Smolensky-RBM-1986, hinton2002training} are an important example of this kind that, moreover, serve as a suitable building block for deeper architectures. Still, quantum annealers have the potential to allow for learning more complex architectures.  

When applying quantum annealing hardware to the learning of Boltzmann machines, the interest is in finding the optimal control parameters that best represent the empirical distribution of a dataset. However, estimating additional shifts for the control parameters, as done by Bian {\it et al.} \cite{bian2010ising}, would not be practical since it is in a sense similar to the very kind of problem that a Boltzmann machine attempts to solve. One could then ask what is the meaning of using a quantum annealer for learning the parameters of a distribution, if to do so we need to use standard techniques to learn the corrections to the control parameters. 

Here we explore a different approach by taking into account only the possibility of an instance-dependent effective temperature without the need to consider further instance-dependent shifts in the control parameters. We devise a technique to estimate the effective temperature associated with a given instance by generating only two sets of samples from the machine and performing a linear regression. The samples used in our effective-temperature estimation algorithm are the same ones used towards achieving the final goal of the sampling application. This is in contrast to the approach taken in Ref. \cite{bian2010ising}, which needs many evaluations of the gradient of the log likelihood of a set of samples from the device, making it impractical for large problem instances. 

We test our ideas in the learning of a special class of restricted Boltzmann machines. In the next section we shall present a brief overview of Boltzmann machines and discuss how quantum annealing hardware can be used to assist their learning. Afterwards, we discuss related work. In the section that follows we introduce our technique to estimate the effective temperature associated with a given instance. We then show an implementation of these ideas for our quantum-assisted learning (QuALe) of a chimera-RBM on the Bars And Stripes (BAS) dataset~\cite{Hinton-BAS-1986, MacKay-book-2002, fischer2012introduction}, implemented in the D-Wave 2X device (DW2X) located at the NASA Ames Research Center. Finally, we present the conclusions of the work and some perspectives of the future work we shall be exploring. 

\section{General considerations}

\subsection{Boltzmann machines }\label{sec:bm}

Consider a binary data set ${\mathcal{D} = \{\bv^1, \dotsc , \bv^D\}}$ whose empiric distribution is $Q(\bv)$; here each datapoint can be represented as an array of Ising variables, i.e. ${\bv^d = (v_1^d,\dotsc ,v_N^d)}$ with ${v_i^d\in\{-1,+1\}}$, for ${i=1,\dotsc , N}$. A Boltzmann machine models the data via a probability distribution ${P(\bv)=\sum_\bu P_B(\bu,\bv)}$, where ${P_B(\bu,\bv) }$ is a Boltzmann distribution on a possibly extended sample space ${\{\bu, \bv\}}$. Here ${\bu = (u_1,\dotsc ,u_M)}$ are the `unobservable' or `hidden' variables, that help capture higher level structure in the data \cite{LeRoux-2008}, and  ${\bv = (v_1,\dotsc ,v_N)}$  are the `visible' variables, that correspond to the data themselves. More precisely, denoting these variables collectively by ${\bs=(\bu,\bv)}$, we can write
\be\label{e:PB}
P_B(\bs)=\frac{e^{-E(\bs)}}{Z},
\ee
where
\be\label{e:E}
E(\bs) = -\sum_{ij\in\mathcal{E}}  W_{ij} s_i s_j - \sum_{i\in\mathcal{V}} b_i s_i 
\ee
is the corresponding energy function, and $Z$ is the normalization constant or partition function. Notice that in this case we do not need a temperature parameter, since it only amounts to a rescaling of the {\it model parameters} $W_{ij}$ and $b_i$ that we want to learn. Here $\mathcal{V}$ and $\mathcal{E}$ are the set of vertices and edges, respectively, that make up the interaction graph $\mathcal{G}=(\mathcal{V}, \mathcal{E})$.

The task is then to find the model parameters that make the model distribution $P$ as close as possible to the data distribution $Q$. This can be accomplished by minimizing the Kullback-Leibler (KL) divergence~\cite{fischer2012introduction}
\be
D_{KL}(Q||P) = \sum_\bv Q(\bv)\ln \frac{Q(\bv)}{P(\bv)},
\ee
between $Q$ and $P$ or, equivalently, by maximizing the average log likelihood 
\be
\mathcal{L}_{\rm av}=\frac{1}{D}\sum_{d = 1}^D\ln P(\bv^d)
\ee
with respect to the model parameters $W_{ij}$ and $b_i$. 

Gradient ascent is a standard method to carry out this optimization via the rule
\BE
W_{ij}^{(t+1)} &=& W_{ij}^{(t)} + \eta \frac{\partial \mathcal{L}_{\rm av}}{\partial W_{ij}} \label{e:rule-W},\\
b_i^{(t+1)} &=& b_i^{(t)} + \eta\frac{\partial \mathcal{L}_{\rm av}}{\partial b_{i}},\label{e:rule-b}
\EE
where $\eta>0$ is the learning rate, and the gradient of the average log-likelihood function is given by \cite{fischer2012introduction}
\BE
\frac{\partial \mathcal{L}_{\rm av}}{\partial W_{ij}} &=&  \bra s_i s_j \ket_\mcD - \bra s_i s_j \ket_\mcM,\label{e:grad_W}\\
\frac{\partial \mathcal{L}_{\rm av}}{\partial b_{i}} &=&  \bra s_i \ket_\mcD - \bra s_i \ket_\mcM .\label{e:grad_b}
\EE
Here $\bra\cdot \ket_\mcD$ denotes the ensemble average with respect to the distribution $P(\bu|\bv)Q(\bv)$ that involves the data. Similarly, $\bra\cdot\ket_\mcM$ denotes the ensemble average with respect to the distribution $P(\bu|\bv)P(\bv) = P_B(\bu,\bv)$ that involves exclusively the model. Such averages can be estimated by standard sampling techniques, such as MCMC. Another possibility, explored in this work, is to rely on a physical process that naturally generates samples from a Boltzmann distribution.

\subsection{Quantum annealing}\label{sec:qa}

Quantum annealing is an algorithm that attempts to exploit quantum effects to find the configurations with the lowest cost of a function describing a problem of interest~\cite{finnila1994quantum,kadowaki_quantum_1998,Farhi2001}. It relies on finding a mapping of such a function into the energy function of an equivalent physical system. The latter is suitably modified to incorporate quantum fluctuations whose purpose is to maintain the system in its lowest-energy solution space. 

In short, the algorithm consists of slowly transforming the ground state of an initial quantum system, which is relatively easy to prepare, into the ground state of a final Hamiltonian that encodes the problem to be solved. The device produced by D-Wave Systems~\cite{Johnson-Nature-2011,Harris-PRB-2010} is a realization of this idea for solving quadratic unconstrained optimization problems on binary variables. It implements the Hamiltonian 
\BE\label{e:H}
H(\tau) &=& A(\tau) H_D + B(\tau) H_P,\\
H_D &=&  -\sum_{i\in \mathcal{V}_C}\sigma_i^x,\\
H_P &=& \sum_{ij\in\mathcal{E}_C} J_{ij}\sigma_i^z\sigma_j^z + \sum_{i\in\mathcal{V}_C} h_i\sigma_i^z,
\EE
where $\sigma_i^{x,z}$ are Pauli matrices that operate on spin or qubit $i$. The {\it control parameters} of the D-Wave machine are composed of a field $h_i$ for each qubit $i$ and a coupling $J_{ij}$ for each pair of interacting qubits $i$ and $j$. The topology of the interactions between qubits in the D-Wave is given by a so-called Chimera graph $\mathcal{C}=(\mathcal{V}_C,\mathcal{E}_C)$. This is made up of elementary cells of $4\times 4$ complete bipartite graphs that are coupled as shown in Fig.~\ref{f:RBM}~(a). The transformation from the simple Hamiltonian $H_D$ to the problem Hamiltonian $H_P$ is controlled by time-dependent monotonic functions $A(\tau)$ and $B(\tau)$, such that $A(0)\gg B(0)$ and $A(1)\ll B(1)$. Here $\tau=t/t_a$, where $t$ is the physical time and $t_a$ is the annealing time, i.e., the time that it takes to transform Hamiltonian $H_D$ into Hamiltonian $H_P$.

Although quantum annealers were designed with the purpose of reaching a ground state of the problem Hamiltonian $H_P$, there are theoretical arguments \cite{Amin-arXiv-2015} and experimental evidence \cite{bian2010ising, PerdomoOrtiz_SciRep2016} suggesting that under certain conditions the device can sample from an approximately Boltzmann distribution at a given effective temperature, as described in more detail in the next section.

\subsection{Quantum annealing for sampling applications}

There are many classical computations that are intrinsically hard and that might benefit from quantum technologies. Common tasks include the factoring of large numbers into its basic primes, as is the case with Shor's algorithm~\cite{Shor1994} in the gate model of quantum computation. Another one described above consists of finding the global minimum of a hard-to-optimize cost function, where quantum annealing is the most natural paradigm. As described at the end of Sec.~\ref{sec:bm}, another computationally hard problem, key for the successful training of Boltzmann machines and related machine learning tasks, is for example the estimation of averages $\bra\cdot\ket_\mcM$ over probability distribution functions $P_B(\bs)$. In the case of models with a slow mixing rate, the standard MCMC approaches would have a hard time obtaining reliable samples from the probability distribution $P_B(\bs)$ \cite{Bengio+Delalleau-2009,Fischer-2010}. As long as the quantum annealer can sample more reliably or more efficiently from this Boltzmann distribution, then we can find value in using it to solve a problem where MCMC might become intractable. It has been pointed out in the literature \cite{LeCun-Nature-2015,Bengio-Book} by several experts in the field that to a large extent the key to success of unsupervised learning relies on breakthroughs in efficient sampling algorithms.

Several key questions arise when considering quantum annealers as potential technologies for providing an algorithmic speed up in sampling applications. Why is a quantum annealer expected to sample from a classical Boltzmann distribution $P_B(\bs)$, given that it is a quantum device? Shouldn't we expect the quantum annealer to sample from a quantum distribution instead? When and why should we expect the quantum annealer to do better than classical MCMC approaches? 

There are several competing dynamical processes happening at different time scales, with the time per annealing cycle being one, and decoherence and relaxation processes having their intrinsic timescale as well. For example, if the annealing time is much larger that the thermal equilibration timescale, the system will remain in its thermal equilibrium until the end of the annealing schedule. On the contrary, if it is too short, diabatic transitions promoting undesirable population flux from the ground state to excited states, would become relevant, leading it to be in a non-equilibrium state.

For quantum annealers that have a strong interaction with the environment leading to relatively fast thermalization and decoherence, theory suggests that the relevant quantum dynamics during an annealing essentially {\it freezes} somewhere between the critical point associated with the minimum gap and the end of the annealing schedule~\cite{Albash_NJP2012,Smelyanskiy_arXiv2015,Amin-arXiv-2015}. In a quasistatic regime~\cite{Smelyanskiy_arXiv2015,Amin-arXiv-2015}, the system happens to be close to a Boltzmann distribution but at a certain effective temperature that is in general different from the physical temperature of the device. Such a {\it freezing point} $\tau_{\rm freeze}$ tends to coincide with the coefficients in Eq.~\eqref{e:H} satisfying $A(\tau_{\rm freeze})\ll B(\tau_{\rm freeze})$, which suggests that the system being quantum annealed might end up in a Boltzmann distribution of the classical cost function encoded in $H_P$.

The intuition behind this phenomenon is that the dominant coupling of the qubits to the environment or bath degrees of freedom is via the $\sigma^z$ operator (for details, see the supplementary material of Refs.~\cite{Johnson-Nature-2011,boixo2014computational}). Since at the freezing point we have $A(\tau_{\rm freeze})\ll B(\tau_{\rm freeze})$, and the interaction with the bath lacks a strong $\sigma^x$ component capable of causing relaxation between the states of the computational basis (i.e., eigenstates of $\sigma^z$), the system cannot relax its population anymore; in other words, its population dynamics freezes. Since around $\tau_{\rm{freeze}}$ the full Hamiltonian driving the dynamics is $H(\tau_{\rm freeze})\approx B(\tau_{\rm{freeze}}) H_P$, if a Boltzmann distribution is indeed reached, it would correspond to an effective temperature $T_{\rm{eff}}$ different from the physical temperature of the device. Here we will follow the convention that the units of temperature are given in a dimensionless energy scale where 1.0 is the maximum programmable value for the $J$ couplers. According to Eq.~\eqref{e:H} the total Hamiltonian at the end of the annealing ($\tau=1$) is given by $H(1)=B(1)H_P$, so $J=1.0$ would correspond to an energy value given by $B(1)$. For the DW2X at NASA, $J=1.0$ corresponds to {$B(1)=7.9$~GHz}. For example, the physical fridge temperature of this quantum annealer, $T_{\text{DW2X}}=12.5$ mK,  corresponds to $T_{\text{DW2X}}=0.033$ in the dimensionless units we follow in this paper. The effective temperature would be $T_{\rm{eff}} \equiv T_{\rm{DW2X}}B(1)/B(\tau_{\rm{freeze}})$; since $B(\tau_{\rm freeze})<B(1)$, then $T_{\rm eff}> T_{\rm{DW2X}}$. Such an effective temperature is expected to depend on the specific instance being studied and on the details of its energy landscape. Some recent unpublished work of our research team indicates that the effective temperature could also be influenced by the noise in the programmable parameters and by its interplay with the specific instance studied, making an {\it a priori} estimation a daunting task. The approach we take in this work is to estimate this effective temperature from the same samples that would be eventually used for the subsequent training process.

We could wonder why a quantum annealer is expected to help in this computational task? It has been shown that quantum tunneling~\cite{boixo2014computational} might be a powerful computational resource for keeping the system close to the ground state and to the proper thermal distribution. It is these quantum resources, available during the quantum dynamics before the freezing point, that might assist and speed up the thermalization process, making sampling more efficient than other classical approaches, such as MCMC. It is important to mention that such a quantum advantage is not expected for all energy landscapes; there will be instances that will be hard for both classical annealers and quantum annealers. The answer to this question will be highly dependent on the quantum resources available and on the complexity of the energy landscape itself. This is an important and interesting question in its own right that we will address in future work. In this work we focus on unveiling the effective temperature that properly defines the distribution we are sampling from and test our method in the context of a machine-learning problem related to the training of Boltzmann machines.

\subsection{Chimera restricted Boltzmann machines} 
\begin{figure}
\includegraphics[width=\columnwidth]{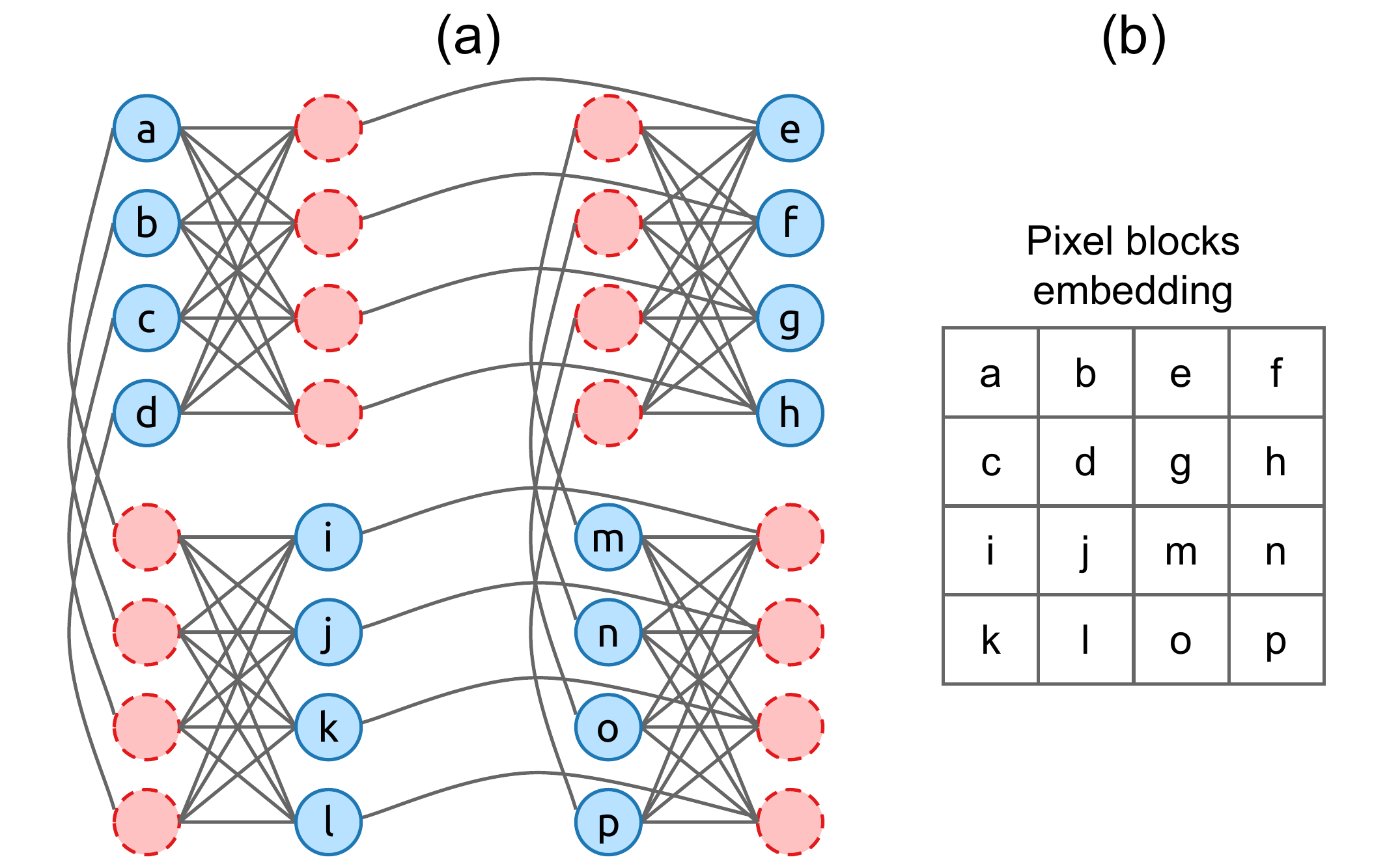}
\caption{ {\it Chimera-RBM and data representation:} (a) D-Wave hardware embedding of a Chimera-RBM with 16 visible and 16 hidden variables. (b) Mapping of the pixels in the pictures to the visible units in the Chimera-RBM that has been used in this work (cf. \cite{Dumolin-2014}). }\label{f:RBM}
\end{figure}

Learning of a Boltzmann machine is, in general, intractable due to the long equilibration time of sampling techniques like MCMC. One way to escape this issue is to use less general architectures. One of the most investigated architectures is the RBM. The interaction graph $\mathcal{G}$ of an RBM is a complete bipartite graph in which visible and hidden units interact with each other, but not among themselves. This implies that the conditional distributions $P(\bv|\bu)$ and $P(\bu|\bv)$ factorize in terms of single variable marginals, which substantially simplifies the problem. One the one hand, data averages $\bra\cdot\ket_\mcD$ can be computed exactly in one shot. On the other hand, model averages $\bra\cdot\ket_\mcM$ can be approximated by $k$-step contrastive divergence (CD-$k$): first, we start with a datapoint $\bv^{(0)}$; then we sample $\bu^{(0)}$ from $p(\bu | \bv^{(0)} )$, and subsequently sample $\bv^{(1)}$ from $p(\bv |\bu^{(0)})$ and so on for $k$ steps. At the end of this process we obtain samples $\bv^{(k)}$ and $\bu^{(k)}$ from which it is possible to estimate model averages~\cite{fischer2012introduction}. CD-$k$ is not guaranteed to give correct results \cite{Bengio+Delalleau-2009, Long-2010} nor does it actually follow the log-likelihood gradient or, indeed, the gradient of any function. Better sampling methods can have, therefore, a positive impact on the kind of models learned. 

It is, in principle, possible to embed an RBM in quantum annealing hardware \cite{Adachi-arXiv-2015}. However, due to limited connectivity of the device, the resulting physical representation would involve a number of qubits and couplings between them much larger than the number of logical variables and weights in the original RBM being represented. It would be preferable to use an alternative model that can be naturally represented in the device. For this reason we will focus on the kinds of models that are obtained after removing from a given RBM all the links that are not present in the D-Wave machine \cite{Dumolin-2014}. We will call this type of model a {\it Chimera Restricted Boltzmann Machine} (Chimera-RBM). Figure~\ref{f:RBM}~-(a) shows an example of a Chimera-RBM, and Fig.~\ref{f:RBM}~(b) shows a possible embedding of the pixels of an image into its visible units (cf. Ref. \cite{Dumolin-2014}).

\section{Related work}
Dumoulin {\it et al.} \cite{Dumolin-2014} have studied the impact of different limitations of quantum annealing hardware for the learning of restricted Boltzmann machines. The authors have focused on three kinds of limitations: noisy parameters, limited parameter range, and restricted architecture. The learning method used was persistent contrastive divergence where the model ensemble averages were estimated with samples from simulated quantum hardware while the data ensemble averages were estimated by exact mean field. 

To assess the impact of limited connectivity, Dumoulin {\it et al.} investigated a Chimera-RBM. They found that limited connectivity is the most relevant limitation in this context. In a sense this is understandable as RBMs are based on complete bipartite graphs while Chimera-RBMs are sparse. Roughly speaking, this means that if the number of variables is of order $N$, the number of parameters present in a Chimera-RBM is a vanishing fraction (of order $1/N$) of the number of parameters in the corresponding RBM. Furthermore, connections in a Chimera-RBM are rather localized. This feature may make capturing higher-level correlations more difficult. 

The authors also found that noise in the parameters of an RBM is the next relevant limitation and that noise in the weights $W_{ij}$ is more relevant than noise in the biases $b_i$. This could happen because the number of biases is a vanishing fraction of the number of weights in an RBM. This argument is no longer valid in a Chimera-RBM, however. The authors also mentioned that noise in the weights changes only when the instance changes, while noise in the biases changes in every sample generated. If this is indeed the case, this could be another reason for the higher relevance of noise in the weights than noise in the biases. 

Finally, an upper bound in the magnitude of the model parameters, similar to the one present in the D-Wave device, does not seem to have much impact. In this respect, we should notice that current D-Wave devices are designed with the sole aim of consistently reaching the ground state. In contrast, typical applications of Boltzmann machines deals with heterogeneous real data which contain a relatively high level of uncertainty, and are expected to exploit a wider range of configurations. This suggests that in sampling applications control parameters are typically smaller than those explored in combinatorial optimization applications. If this is indeed the case, potential lower bounds in the magnitude of the control parameters can turn out to be more relevant for sampling applications. In this respect, it is important to notice that noise in the control parameters can lead to an effective lower bound. 

While Dumoulin {\it et al.} modeled the instance-dependent corrections as independent Gaussian noise around the user defined parameter values, Denil and De Freitas \cite{Denil-2011} devised a way to by-pass this problem altogether. To do this, the authors optimized the one-step reconstruction error as a black-box function and approximated its gradient empirically. They did this using a technique called simultaneous perturbation stochastic approximation. However, with this approach, it is not possible to decouple the model from the machine. Furthermore, it is not clear what the efficiency of this technique is or how to extend it to deal with the more robust log-likelihood function instead of the reconstruction error. In their approach only the hidden layer is embedded in the D-Wave, and qubit interactions are exploited to build a semi-restricted Boltzmann machine. Although they reported encouraging results, the authors acknowledged that they are still not conclusive.

More recently, Adachi and Henderson \cite{Adachi-arXiv-2015} devised a way to embed an RBM on a D-Wave chip with Chimera topology. They did this by representing each logical variable by a string of qubits with strong ferromagnetic interactions. Furthermore, they implemented a simple strategy to average out the effects of the noise in the D-Wave control parameters. They used the quantum annealer to estimate model averages as in Ref. \cite{Dumolin-2014} for pre-training a two-layer neural network. However, the authors did not evaluate the performance of the quantum device at this stage; they rather post trained the model with (classical) discriminative techniques for learning the labels of a coarse-grained version of a well-known data set of handwritten digits called MNIST and computed the classification error. They reported that this approach outperforms the standard approach where CD-1, instead of quantum annealing, is used for pre training the generative model. 

\section{Quantum-assisted learning of Boltzmann machines}

In this work we assume that quantum annealers, like those produced by D-Wave Systems, sample from a Boltzmann distribution defined by an energy function as in Eq. \eqref{e:E}, with $W_{ij}=J_{ij}/T_{\rm eff}$ and $b_i = h_i/T_{\rm eff}$, where $T_{\rm eff}$ can be instance-dependent. While the control parameters for the D-Wave are couplings and fields, i.e. $J_{ij}$ and $h_i$, the learning takes place on the ratio of the control parameters to the temperature, i.e. $W_{ij}$ and $b_i$. Inferring temperature is therefore a fundamental step to be able to use samples from a device like D-Wave for learning, since it provides a translation from $\{W_{ij}\}$ to $\{J_{ij}\}$ and from $\{b_i\}$ to $\{h_i\}$. We propose a QuALe technique that includes an efficient estimation of the effective temperature. It is initialized as follows: 
\begin{itemize}
\item Pick small initial control parameters $J_{ij}^{(0)}$~and~$h_i^{(0)}$, and sample from the device. 
\item Using the samples obtained in the previous item, estimate the initial temperature $T_{\rm eff}^{(0)}$ to compute the initial model parameters ${W_{ij}^{(0)} = J_{ij}^{(0)}/T_{\rm eff}^{(0)}}$ and ${b_i^{(0)} = h_i^{(0)}/T_{\rm eff}^{(0)}}$.
\end{itemize}

Then the technique iterates as follows: 
\begin{itemize}
\item Using the samples and model parameters obtained in step $t$, estimate the corresponding temperature $T_{\rm eff}^{(t)}$ and update the model parameters according to Eqs. \eqref{e:rule-W} and \eqref{e:rule-b} to obtain $W_{ij}^{(t+1)}$ and $b_i^{(t+1)}$.
\item Obtain new control parameters by performing ${J_{ij}^{(t+1)}\approx T_{\rm eff}^{(t)}W_{ij}^{(t+1)}}$ and ${h_i^{(t+1)}\approx T_{\rm eff}^{(t)} b_i^{(t+1)}}$ and sample from the device. 
\end{itemize}
A few comments are in order. First, for each sample step we need to generate samples for estimating model and data ensemble averages. For the former we just need to run the device with the specified control parameters. For the latter we need to generate samples with the visible units clamped to the data points, which can be done by applying suitable fields to the corresponding qubits. However, in the case of restricted Boltzmann machines we can avoid this last step as it is possible to compute exactly the data ensemble averages.
Second, notice that to compute the new control parameters in step $t+1$ it would have been ideal to estimate the temperature $T_{\rm eff}^{(t+1)}$ in the same step. However, to estimate such a temperature we would need to know what the parameters are at time $t+1$. To escape this vicious cycle we have set $T_{\rm eff}^{(t+1)}\approx T_{\rm eff}^{(t)}$. Finally, notice that if we think of the learning process in terms of the control parameters $J_{ij}$ and $h_i$, we may get the impression that the learning rate is temperature-dependent. We would like to emphasize that the learning operates on the model parameters $W_{ij}$ and $b_i$, which are the one that actually shape the Boltzmann distribution through the update rules given by Eqs. \eqref{e:rule-W} and \eqref{e:rule-b}. So, the actual learning rate is given by $\eta$ in the update equations above; if we fix $\eta$ to a constant, it would remain so. We need $T_{\rm eff}$ only to estimate the required control parameters. Still, the approximation $T_{\rm eff}^{(t+1)}\approx T_{\rm eff}^{(t)}$ and the error in their estimation can introduce noise that cause the learning process to deviate from the actual update rules given by Eqs. \eqref{e:rule-W} and \eqref{e:rule-b}. It would be interesting to investigate what the impact of this noise is in contrast to that due to the estimation of the log-likelihood gradient with a finite number of samples. In the next section we discuss a method for estimating this instance-dependent temperature.

\section{Temperature estimation}\label{s:T}
\subsection{Extracting temperature from two sample sets}\label{s:log}
At a generic inverse temperature $\beta$, the probability of observing a configuration of energy $E$ is given by ${P_\beta(E) = g(E) e^{-\beta E}/Z(\beta)}$ . Here $g(E)$ is the degeneracy of the energy level $E$ and the normalization factor, $Z(\beta)$, is the partition function. We want to devise an efficient method for estimating the effective temperature associated with a given instance. To do this, consider the log-ratio of probabilities associated with two different energy levels, $E_1$ and $E_2$, given by
\be\label{e:l}
\ell (\beta) \equiv \ln \frac{P_\beta(E_1)}{P_\beta(E_2)} = \ln \frac{g(E_1)}{g(E_2)} - \beta\Delta E,
\ee
where $\Delta E = E_1-E_2$. We can estimate this log-ratio by estimating the frequencies of the two energy levels involved; in practice, we may have to do a suitable binning to have more robust statistics. Although we cannot control the physical temperature, we could in principle do this for different values of the parameter $\beta$ by rescaling the control parameters of the device by a factor $x$. This is equivalent to setting a parameter $\beta = x\beta_{\rm eff}$, where $\beta_{\rm eff} = 1/T_{\rm eff}$ is the inverse of the effective temperature $T_{\rm eff}$ associated with the instance of interest. Notice that this is only true under the assumption that $T_{\rm eff}$, despite being generally dependent on arbitrary variations of the control parameters, does not change appreciably under these small rescalings. By plotting the log-ratio $\ell(x\beta_{\rm eff})$ against the scaling parameter $x$, we should obtain a straight line whose slope and intercept are given by $-\beta_{\rm eff}\Delta E$ and $\ln [g(E_1)/g(E_2)]$, respectively. Since we know the energy levels we can in principle infer $\beta_{\rm eff}$. However, the performance of this method was rather poor in all experiments we carried out (not shown). A reason could be that to perform the linear regression and extract the corresponding effective temperature, several values of $x$ need to be explored in a relatively wide range. Next we present a proposal that mitigates this limitation, which also happens to be much more efficient.

The previous approach relied on several values of the scaling parameter $x$ but only two energy levels. We were not exploiting all the information available in the other energy levels sampled from the quantum annealer. We can exploit such information to obtain a more robust estimate of the temperature by sampling only for the original control parameters and a single rescaling of them. The idea is to take the difference $\Delta\ell\equiv \ell(\beta)-\ell(\beta^\prime)$, with $\beta = \beta_{\rm eff}$ and $\beta^\prime =x\beta_{\rm eff}$, to eliminate the unknown degeneracies altogether, yielding
\be\label{e:dl}
\Delta\ell = \ln \frac{P_\beta(E_1)P_{\beta^\prime}(E_2)}{P_\beta(E_2)P_{\beta^\prime}(E_1)} = \Delta\beta\Delta E ,
\ee
where $\Delta \beta = \beta^\prime - \beta = (x-1)\beta_{\rm eff}$. In this way, by generating a second set of samples at a suitable value of $x$ and then taking the differences of all pairs of populated levels, we can plot $\Delta\ell$ against $\Delta E$. According to Eq. \eqref{e:dl} this is expected to be a straight line with slope given by $(x-1)\beta_{\rm eff}$. In practice, one has to choose a binning strategy and use the same bin intervals in both histograms so that the \textit{overlap} makes sense. For example, by setting the number of bins to $K=\left\lceil \sqrt{2R} \right\rceil$, where $R$ is the number of samples per set, one obtains $\mathcal{O}(K^2)=\mathcal{O}(R)$ data points for linear regression.
Notice that the raw energies computed before binning refer to the {\it original} values of the control parameters in {\it both} cases, {\it not} the rescaled ones. This is because we have already counted the effect of the rescaling in a different inverse effective temperature $\beta^\prime = x\beta_{\rm eff}$. Finally, the energy levels obtained after binning correspond to the midpoint of each bin.

The choice of $x$ matters: if it is too small no informative changes would be detected, other than noise due to finite sampling and uncontrolled physical processes in the device. If it is too large, several levels would become unpopulated and we would not be able to compare them at both the original and rescaled control parameters; moreover, the assumption of the invariance of $T_{\rm eff}$ under small perturbations around the original control parameters would be less likely to be valid. Next, we discuss how to choose the value of $x$.

\subsection{A rule of thumb for the scaling factor}
We can rely on concepts of information theory to guide the choice of the scaling factor $x$. The idea here is to choose the value of $x$ as close as possible to one that still allows us to distinguish between the two sets of samples of a given size. Via Sanov's theorem, the KL divergence provides a natural way to characterize the notion of distinguishability in this case \cite{Balasubramanian-1997, Myung-PNAS-2000, Mastromatteo-PhD-2013}. Here we will briefly discuss the main ideas in a rather informal way; the interested reader can refer to Ref. \cite{Mastromatteo-PhD-2013} for details. We want to know whether we can distinguish between two Boltzmann distributions at different inverse temperatures $\beta$ and $\beta^\prime$ from a set of $R$ samples. To do this, it is useful to consider that we compute the maximum-likelihood estimate of the inverse temperature $\beta_{\rm ML}$ from the sample set corresponding to inverse temperature $\beta$. We can consider that we repeat this procedure many times so we can compute the probability distribution of $\beta_{\rm ML}$. The two Boltzmann distributions are said to be {\em distinguishable} from a set of $R$ samples if the probability of $\beta_{\rm ML}$ being close to $\beta^\prime$ is smaller than a given tolerance $P_0$, i.e., if
\be
{\rm Prob}\left[|\beta_{\rm ML} - \beta^\prime|<\delta\right]< P_0 ,
\ee
where $\delta$ is a suitably small constant. From Sanov's theorem it follows that when $R$ is large enough
\be\label{e:Sanov}
{\rm Prob}\left[|\beta_{\rm ML} - \beta^\prime|<\delta\right]\approx C e^{-R D_{\rm KL}(P_{\beta^\prime}||P_\beta)},
\ee
where the factor $C$ gathers sub-dominant terms in $R$. So, if ${D_{\rm KL}(P_{\beta^\prime}||P_\beta) >  \ln (C/P_0) / R}$ the two Boltzmann distributions are distinguishable in the sense defined above.

Assuming that $\beta$ and $\beta^\prime$ are close enough, the KL divergence can be expanded up to second order to yield
\be\label{e:KLsmall}
D_{\rm KL}(P_{\beta^\prime}||P_\beta) \approx \frac{1}{2}\chi(\beta) \Delta\beta^2, 
\ee 
where 
\be \label{e:FI}
\chi(\beta) = \frac{\partial^2\ln Z(\beta)}{\partial\beta^2} = \bra E^2\ket - \bra E\ket^2 \equiv \sigma_E^2
\ee
is known in information theory as the Fisher information, or generalized susceptibility; in this case, it is essentially the specific heat. When $R$ is large enough, the right hand side in Eq. \eqref{e:Sanov} becomes appreciable only for very close $\beta$ and $\beta^\prime$. So for large $R$ we can replace the KL divergence by the Fisher information in Eq. \eqref{e:Sanov}.

Following these ideas, we propose to choose the scaling factor $x$ such that ${\frac{1}{2}\chi(\beta) (1-x)^2\beta_{\rm eff}^2 = d_{\rm KL}/R}$, where $d_{\rm KL}$ is a given constant (cf. Ref. \cite{Habeck-arXiv-2015}). Eqs.~\eqref{e:KLsmall}~and~\eqref{e:FI} yield 
\be\label{e:x}
{x = 1\pm \sqrt{\frac{2\, d_{\rm KL}}{R\, \beta_{\rm eff}^2 \, \sigma_E^2}}.}
\ee
Some remarks are in order: (i) Eq. \eqref{e:x} gives a rule of thumb to choose a suitable value of $x$ for estimating $\beta_{\rm eff}$; however, the latter also appears in this expression. We can initiate $\beta_{\rm eff}$ by either making a reasonable guess or using the pseudo-likelihood estimate (see the Appendix \ref{s:PLE}). (ii) The sign in Eq. \eqref{e:x} could be chosen to be positive during the first iterations to avoid the rescaled control parameters being below the noise level of the device, and negative afterwards to avoid the rescaled control parameters being above the allowed range. (iii) Equation~\eqref{e:x} has been derived assuming that values of the KL divergence about $d_{\rm KL}/R$ can be well approximated with the Fisher information. These assumption may fail in practice when $R$ is relatively small or when $x$ is far from the reference value at $x=1.0$. (iv) In principle, as long as the samples generated by the quantum annealer follow a Boltzmann distribution and the effective temperature remains constant under re-scalings of the control parameters, our temperature estimation technique is exact if there are enough samples. Still, the number of samples needed could grow exponentially with problem size due to the bias and variance associated with our estimator, whose study we leave for future work. (v) Finally, the linear regression to compute our estimator may be affected by noise due to energy bands with very low frequency; in principle, this could be mitigated by relying on a weighted linear regression giving more weight to points associated with higher frequencies. 

\section{A few gadgets to improve performance}

In this section we discuss three techniques that help improve the performance of our quantum-assisted learning algorithm. First of all, it is known that the performance of quantum annealers can be significantly impaired by the presence of both persistent and random biases between the actual values of the control parameters and the user-specified values. Perdomo-Ortiz {\it et al.}  \cite{PerdomoOrtiz_SciRep2016} have developed a technique for determining and correcting the persistent biases and have shown evidence that this recalibration procedure can enhance the performance of the device for solving combinatorial optimization problems. In the next section we will show evidence that correcting for persistent biases can also enhance the performance of quantum annealers for sampling applications. 

Second, noise in the control parameters can hinder the initial stage of learning when they are typically small. In order to avoid this situation we can run CD-$1$ for a few iterations until we find meaningful initial values for the control parameters that are above the noise level of the device and then restart with QuALe. This is exclusively due to the current state of quantum annealing technologies and it is expected to be further mitigated in new generations of these devices. We emphasize that the number of iterations with CD-1 has to be small to keep the weights within the dynamical range of the device.

Finally, to estimate the effective temperature associated with a given instance we need to generate two sets of samples: one corresponding to the actual values of the control parameters that we are interested in, and another corresponding to these values rescaled by a factor $x$. According to the discussion in the previous section, the scaling factor is chosen in such a way that the two probability distributions are as close as possible, yet distinguishable. So, we expect that the samples obtained at $\beta^\prime = x\beta_{\rm eff}$ can also be used for the estimation of the log-likelihood gradient, given by Eqs. \eqref{e:grad_W} and \eqref{e:grad_b}, at $\beta = \beta_{\rm eff}$ via the technique of importance sampling \cite{Bishop-book-2006}. In short, we can use a set of samples $\{\bs^1,\dotsc , \bs^R\}$ extracted from a Boltzmann distribution at inverse temperature $\beta^\prime$ to estimate ensamble averages of an arbitrary observable $A$ with a Boltzmann distribution at inverse temperature $\beta$ as
\be\label{e:IS}
\bra A\ket_\beta \approx \frac{\sum_{r=1}^R \rho(\bs^r) A(\bs^r)}{\sum_{r=1}^R \rho(\bs^r)},
\ee
where $\rho(\bs) = e^{-(\beta - \beta^\prime) E(\bs)}$ is the ratio between the unnormalized probabilities. The approximation is expected to be good as long as the two distributions are close enough \cite{Bishop-book-2006}.
In the next section we will show evidence that including the set of samples corresponding to the rescaled control parameters indeed improves the performance of QuALe. 

From now on, when refering to the QuALe algorithm we imply that these three gadgets are also included, unless otherwise specified. 

\section{Learning of a Boltzmann machine assisted by the D-Wave 2X}

Now that we have at our disposal a robust temperature estimation technique, we can use it for learning Boltzmann machines. We decided to focus on the learning of a Chimera-RBM for two reasons. On the one hand, although an RBM can be embedded into quantum hardware \cite{Adachi-arXiv-2015}, it requires us to represent single variables with chains of qubits coupled via ferromagnetic interactions of a given strength. Instead of forcing couplings to take a specific value to meet a preconceived design, it might be better to allow the learning algorithm itself to find the parameter values that work best for a particular application. On the other hand, the focus of our work is on better understanding the challenges that need to be overcome for using quantum annealers for sampling applications and taking the necessary steps towards an effective implementation of deep learning applications on quantum annealers.

This systematic study provides both an assessment of the use of the D-Wave in learning Boltzmann machines and a study of the impact of the effective temperature in the learning performance. We consider it important to assess the performance of the different methods by computing the exact log-likelihood during the learning process. Otherwise, we could not be sure whether a difference in performance is due to the new learning method or due to errors in the approximation of the log-likelihood. For this reason we tested the method on a small synthetic data set called Bars and Stripes (BAS) and computed exhaustively the corresponding log-likelihood for evaluation. The BAS dataset consists of $4\times 4$ pictures generated by setting the four pixels of each row (or column) to either black (-1) or white (+1), at random \cite{Hinton-BAS-1986, MacKay-book-2002,fischer2012introduction}. Another reason to focus on this small synthetic dataset is that while generating, e.g., 2000 samples in the DW2X for a given instance can take about 40 ms, the waiting time for accessing the machine to generate a new set of samples for a different instance can vary widely depending on the amount of jobs that are scheduled. So, while running QuALe with 2000 samples per iteration on the whole chip (1097 qubits) for $10^4$ iterations could take in principle about 7 min if we had exclusive access to the device, the waiting times of the different jobs could increase this time by several orders of magnitude. 

\begin{figure}
\includegraphics[width=\columnwidth]{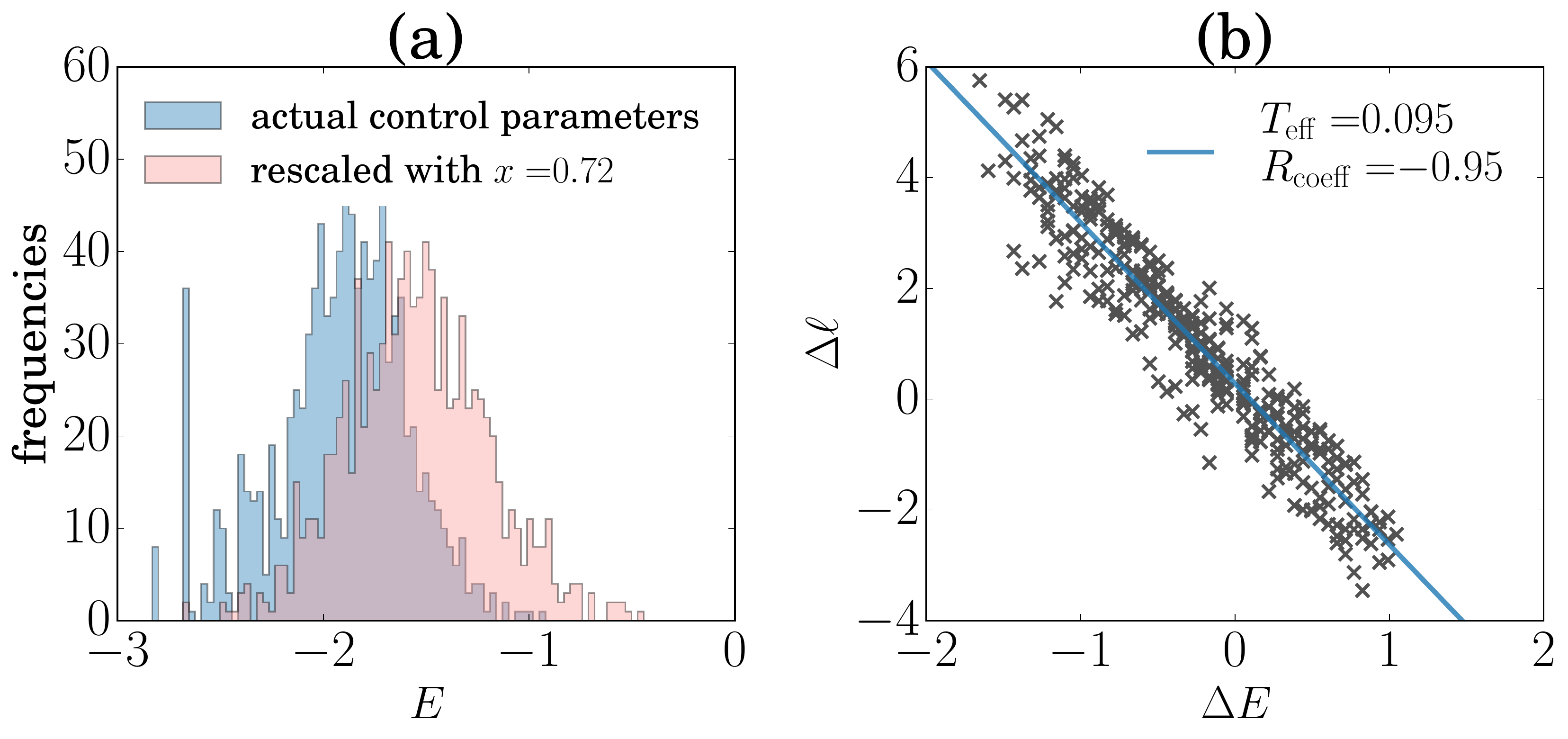}
\caption{{\it  Temperature estimation:}  (a) Energy histograms obtained from $R=1000$ samples generated by the DW2X for two different sets of control parameters and $K=\left\lceil \sqrt{2R} \right\rceil$ bins. The (blue) histogram that is shifted to the left corresponds to a typical set of control parameters found during the learning of a Chimera-RBM on the BAS dataset (cf. Fig. \ref{f:train}) using $d_{KL}/R =1/2$. The (pink) histogram shifted to the right corresponds to these control parameters scaled by a factor $x=0.72$ obtained using Eq. \eqref{e:x}.  (b) Log-likelihood ratio differences $\Delta\ell$ in Eq. \eqref{e:dl} plotted against the corresponding energy differences $\Delta E$. These are computed using all energy bins in the overlap of the two histograms in (a). The straight line is obtained by a linear regression using least squares and predicts, according to Eq. \eqref{e:dl}, an effective temperature of $T_{\rm eff}\approx 0.095$ and a regression coefficient of $R_{\rm coeff}\approx -0.95$. The units of temperature are given in a dimensionless energy scale where 1.0 is the maximum programmable value for the $J$ couplers. For the DW2X at NASA, $J=1.0$ corresponds to 7.9 GHz. For example, the physical fridge temperature of this quantum annealer, $T_{\text{DW2X}}=12.5$ mK,  corresponds to $T_{\text{DW2X}}=0.033$ in the dimensionless units we follow in this paper. The instance plotted here has a $T_{\rm{eff}} \approx 3 T_{\text{DW2X}}$}  \label{f:T}
\end{figure}

\begin{figure}
\includegraphics[width=\columnwidth]{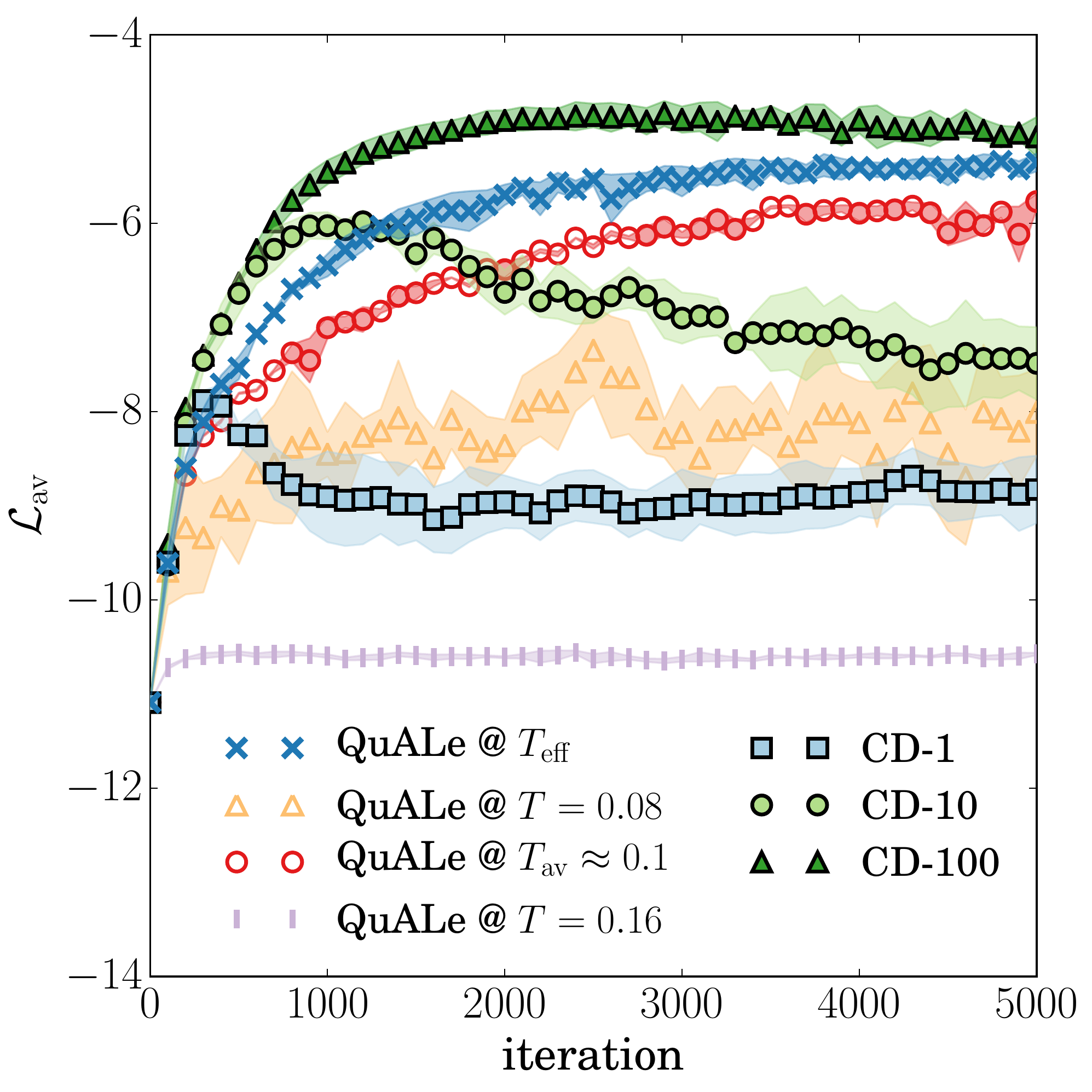}%
\caption{{\it Comparison of learning algorithms:} Average performance of different algorithms for the learning of a Chimera-RBM on the $4\times 4$ BAS dataset. The performance is measured in terms of the average log-likelihood $\mathcal{L}_{\rm av}$, which has been evaluated exhaustively every 50 iterations. All points correspond to average of $\mathcal{L}_{\rm av}$ over five different runs on the same location in the DW2X chip, and the bands correspond to one standard deviation. The (blue) diagonal crosses correspond to quantum-assisted learning estimating effective temperatures (QuALe@$T_{\rm eff}$) with the DW2X using $R=1000$ samples in each iteration for the estimation of both log-likelihood gradient and temperature for actual and rescaled control parameters. The (red) empty circles correspond to fixed-temperature quantum-assisted learning algorithm (QuALe@$T_{\rm av}\approx 0.1$), using the average temperature $T_{\rm av}\approx 0.1$ found during the run of QuALe@$T_{\rm eff}$. The vertical lines and empty triangles correspond to fixed-temperature quantum-assisted learning algorithm using temperatures above and below the average temperature $T_{\rm av}$, namely $T = 0.16$ (QuALe@$T=0.16$) and $T=0.08$ (QuALe@$T=0.08$). The filled squares, circles, and triangles correspond to learning using CD-$k$ for $k = 1,\, 10,\, 100$, respectively. 
} \label{f:train}
\end{figure}

We modeled the BAS dataset with a Chimera-RBM of 16 visible and 16 hidden units with the topology shown in Fig. \ref{f:RBM} (a). The mapping of pixels to visible units is shown in Fig. \ref{f:RBM} (b) (cf. \cite{Dumolin-2014}). We run all algorithms with learning rate $\eta = 0.03$, which is the best value we found among five values in the range $[0.01, 0.1]$. To begin with, Fig. \ref{f:T} shows an instance of temperature estimation using $R=1000$ samples from the DW2X and $d_{\rm KL}=500$, for generic control parameters found during the learning process (see Fig. \ref{f:train}). This value of $d_{\rm KL}$ is the one that worked best out of a few trial values.  Fig.~\ref{f:T}a shows the histograms corresponding to $K=\left\lceil \sqrt{2R} \right\rceil$ bins of samples obtained at the actual control parameters (blue histogram, shifted to the left) and the rescaled ones (pink histogram, shifted to the right). Fig. \ref{f:T}b shows a plot of $\Delta\ell$ against $\Delta E$ for all energy values that appear in the overlap of the two histograms. We can observe a rather clear linear trend as predicted by Eq. \eqref{e:dl}, which is confirmed by a relatively high regression coefficient, $R_{\rm coeff} \approx -0.95$. From the slope $m$ of the regression line we can obtain the effective temperature by solving $m=\Delta\beta = (x-1)\beta_{\rm eff}$.

\begin{figure}
  \includegraphics[width=\columnwidth]{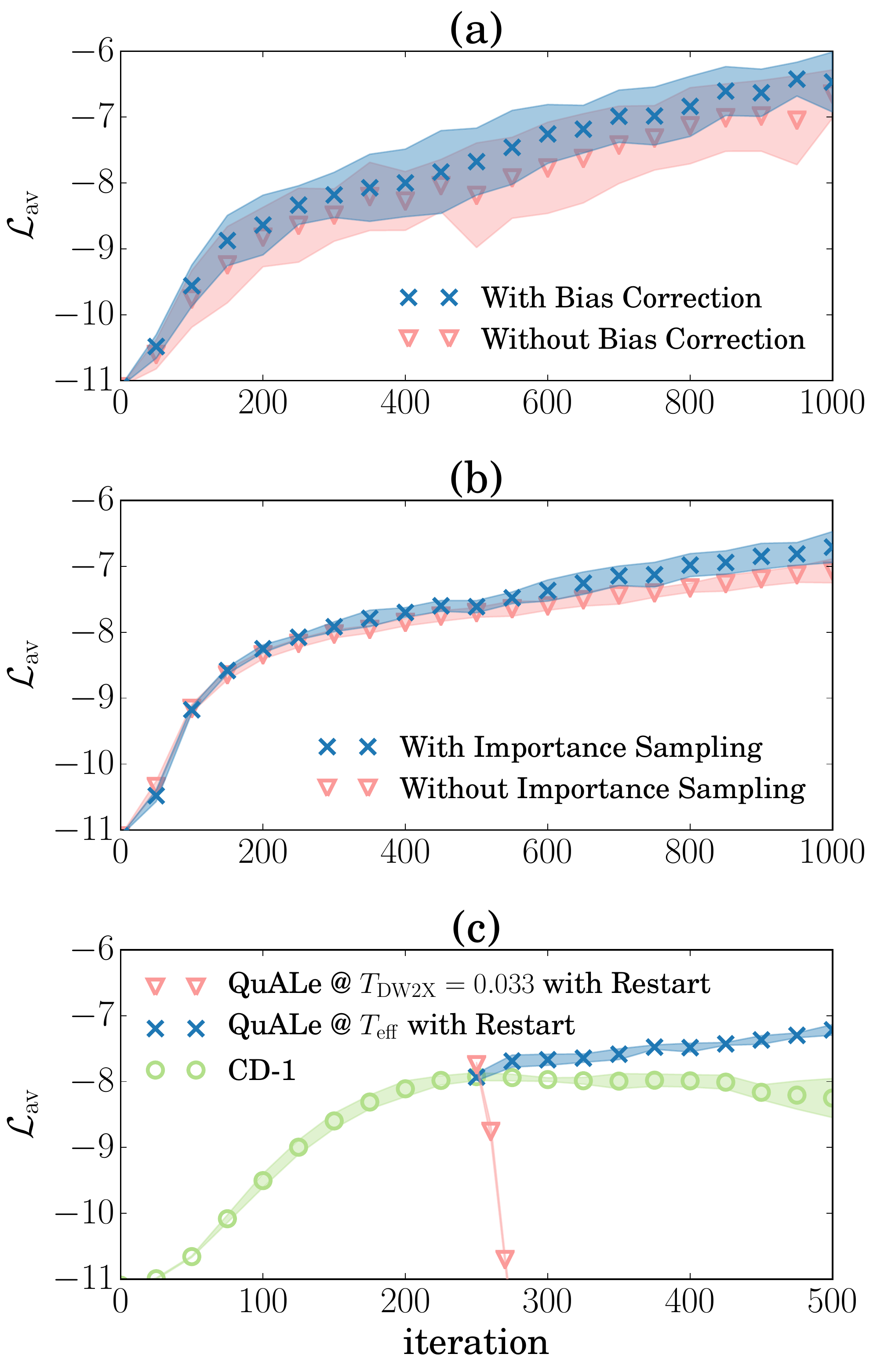}
\caption{{\it Impact of added gadgets:} Average performance of the quantum-assisted learning of a Chimera-RBM on the $4\times 4$ BAS dataset. The performance is measured in terms of the average log-likelihood $\mathcal{L}_{\rm av}$, which has been evaluated exhaustively every 50 iterations. (a) QuALe@$T_{\rm eff}$ with (blue crosses) and without (pink triangles) persistent bias correction. These results are obtained by implementing a Chimera-RBM on five different locations of the DW2X chip and running the QuALe algorithm three times on each location, for a total of 15 runs. The points correspond to the average of $\mathcal{L}_{\rm av}$ over those 15 runs and the bands to one standard deviation.  (b) QuALe@$T_{\rm eff}$ with (blue crosses) and without (pink triangles) taking into account the samples obtained at $x\neq 1$ for the estimation of the log-likelihood gradient, via importance sampling. The points correspond to the average of $\mathcal{L}_{\rm av}$ over five runs of QuALe on a single location of the DW2X chip. (c) QuALe@$T_{\rm eff}$ (blue crosses) starting after a given number of iterations of CD-1 to escape the noise level of the DW2X. Each point represents the average of $\mathcal{L}_{\rm av}$ over five runs of each algorithm and the error bars correspond to one standard deviation.
Notice the dramatic drop in performance of a naive suboptimal version of QuALe@$T_{\rm{DW2X}}$ that uses the physical temperature instead of estimating $T_{\rm{eff}}$ as suggested in this work. The value for QuALe@$T_{\rm{DW2X}}$ is out of the range of the plot and oscillates around $\mathcal{L}_{\rm av} = -14$.} \label{f:gadgets}
\end{figure}

Fig. \ref{f:gadgets}a shows the impact of bias correction on the performance of the QuALe algorithm. The performance is measured in terms of the average log-likelihood $\mathcal{L}_{\rm av}$, which has been evaluated exhaustively every 50 iterations. These results are obtained by implementing the Chimera-RBM on five different locations of the DW2X chip and running the QuALe algorithm three times on each location, for a total of 15 runs. The points correspond to the average of $\mathcal{L}_{\rm av}$ over those 15 runs and the error bars correspond to one standard deviation. We can see that QuALe with persistent bias correction (blue crosses) outperforms QuALe without it (pink triangles). Fig \ref{f:gadgets}b, on the other hand, shows the QuALe algorithm with (blue crosses) and without (pink triangles) taking into account the samples obtained at $x\neq 1$ for the estimation of the log-likelihood gradient, via importance sampling. The points correspond to the average of $\mathcal{L}_{\rm av}$ over five runs of QuALe on a single location of the DW2X chip. Finally, Fig. \ref{f:gadgets}c shows the positive impact of carrying out a few iterations of CD-1 to generate suitable initial conditions for QuALe. 

Fig. \ref{f:train} shows the evolution of $\mathcal{L}_{\rm av}$ during the learning of a Chimera-RBM on the BAS dataset under different learning algorithms, all of them with learning rate $\eta=0.03$. We can observe that the quantum assisted learning algorithm with effective-temperature estimation at each iteration (QuALe@$T_{\rm eff}$, blue diagonal crosses) outperforms {CD-$1$} (blue solid squares) after about 300 iterations and CD-10 (green solid circles) after about 1500 iterations. However, within the 5000 iterations shown in the figure, QuALe@$T_{\rm eff}$ has not yet been able to outperform CD-100, although there is a clear trend in that direction. As we did not observe any significant improvement when using larger values of $k$, we expect that {CD-$100$} is close to an exact computation (cf. Theorem 5.1 in~\cite{Bengio-book-2009}). Interestingly, all CD-$k$ reach their best average performance after a relatively small number of iterations while QuALe@$T_{\rm eff}$, in contrast, increases slowly and steadily. One may be inclined to think this is because CD-$k$ estimates the model averages from samples generated by a $k$-step Markov chain initialized at each data point. In this way CD-$k$ is using information contained in the data from the very beginning for the estimation of the model ensemble averages, while QuALe@$T_{\rm eff}$ ignores them altogether. However, if this were indeed the case one should expect such a trend to diminish for increasing values of $k$, something that is not observed in the figure. A better understanding of this point has the potential to considerably improve the performance of QuALe@$T_{\rm eff}$.

To assess the relevance of temperature estimation for QuALe@$T_{\rm eff}$, we also show in Fig. \ref{f:train} the average performance under quantum assisted learning at a fixed temperature. First, it is worth mentioning that using the physical temperature of the device, $T_{\rm DW2X}=0.033$ (corresponding to $T_{\rm DW2X} = 12.5$ mK as explained in the caption of Fig.~\ref{f:gadgets}), leads to a very poor performance, reaching values $\mathcal{L}_{\rm av} < -14$ (not shown). Fixing the temperature to the average  QuALe@$T_{\rm av} \approx 0.1$ over all temperatures found during the run of QuALe@$T_{\rm eff}$ leads to a better performance (red empty circles), but still well below that displayed by QuALe@$T_{\rm eff}$ itself. Fixing the temperature to $T_0=0.08 < T_{\rm av}$ (QuALe@$T=0.08$) and to $T_0=0.16 > T_{\rm av}$ (QuALe@$T=0.16$) leads to a decrease in performance with respect to that displayed with $T_{\rm av}$.

\begin{figure}
\includegraphics[width=\columnwidth]{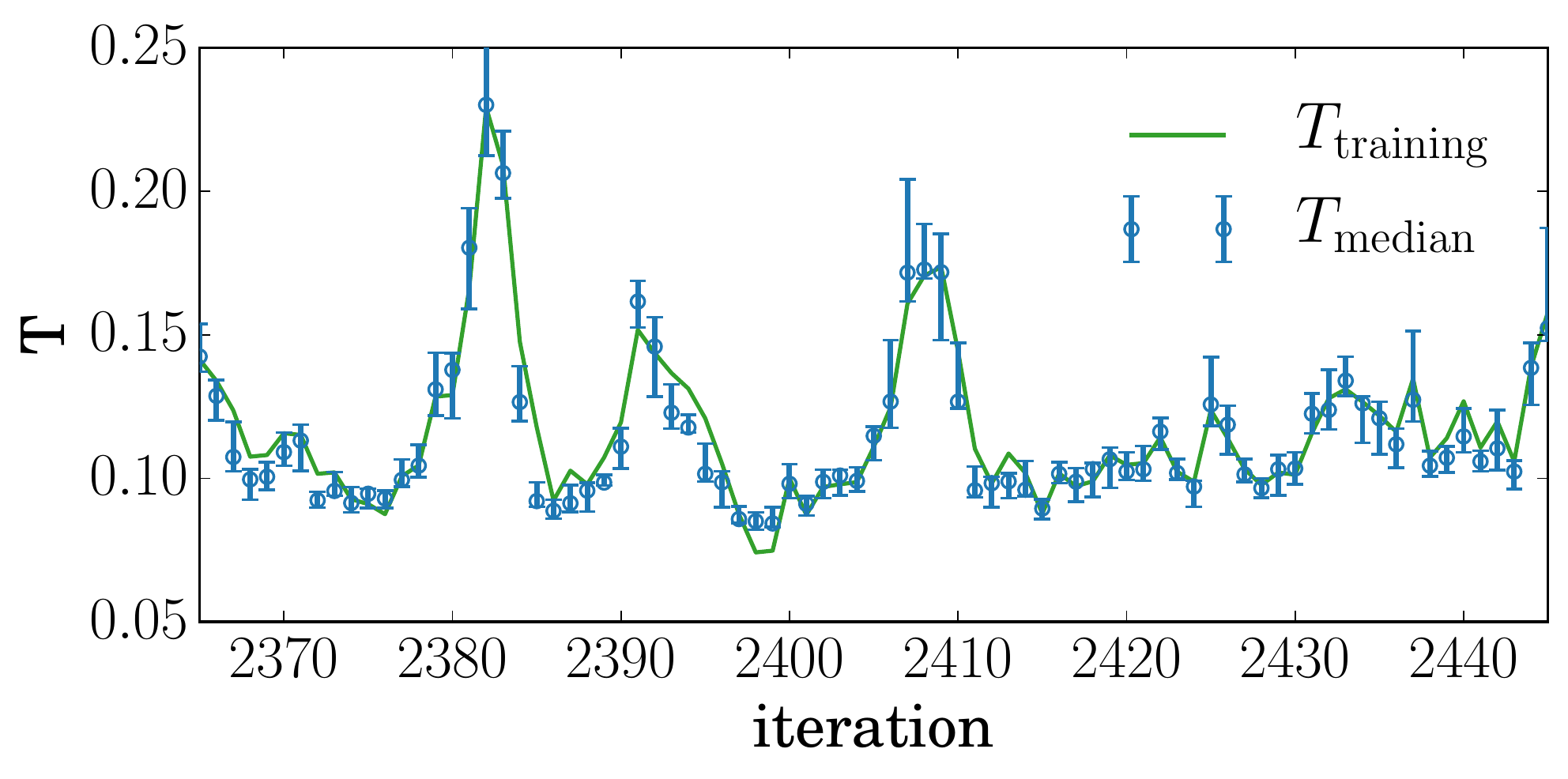}
\caption{{\it Variation of effective temperatures during learning:} The (green) line shows the values of the effective temperature during 80 iterations of QuALe@$T_{\rm eff}$ on the BAS dataset, starting from iteration 2365. The (blue) circles correspond to the median of the effective temperature estimated 15 times for each instance of the control parameters found during a learning session. The error bars represent the first and third quartiles.  
} \label{f:T_QuALe}
\end{figure}

In Fig. \ref{f:T_QuALe} we can observe the variation of the effective temperature estimated during a window of 80 iterations of QuALe@$T_{\rm eff}$ (green line). To evaluate whether such a variation is within the finite sampling error, we estimated the effective temperature 15 times at each iteration. The blue circles show the median of $T_{\rm eff}$ and the error bars represent the corresponding first and third quartiles. Clearly, this variation cannot be explained as due to finite sampling error. We emphasize that during the execution of QuALe the effective temperature is estimated only once.

\section{Conclusions and Future Work}

Applications that rely on sampling, such as learning Boltzmann machines, are in general intractable due to long equilibration times of sampling techniques like MCMC~\cite{Sinclair-InfComp-1989,Frigessi-Biometrika-1997}. Some authors have conjectured quantum annealing could have an advantage in sampling applications. In this work we proposed a strategy to overcome one of the main limitations when intending to use a quantum annealer to sample from Boltzmann distributions: the determination of effective temperatures. The simple technique proposed in this work uses samples obtained from a quantum annealer (the DW2X at NASA is our experimental implementation) to estimate both the effective temperature and the hard-to-compute term in the log-likelihood gradient, i.e., the averages over the model distribution; these are needed to determine the next step in the learning process. We present a systematic study of the impact of the effective-temperature in the learning of a Chimera-RBM model with 16 visible and 16 hidden units. To do so, we compared the QuALe algorithm with both instance-dependent effective temperature and different constant effective temperatures to the performance of a CD-$k$ implementation, with $k$ equal to $1$, $10$, and $100$.

The Chimera-RBM model itself is much less powerful than the RBM model. While the former is sparse with a number of parameters increasing linearly with the number of variables, the latter is dense with a number of parameters increasing quadratically. For instance, the Chimera-RBM that we have studied here, with 16 hidden and 16 visible variables, has only about 31\% of the weight parameters that a corresponding RBM of the same size has. This is reflected by the fact that a Chimera-RBM, trained either with QuALe or with standard classical techniques, struggles to generate samples faithfully resembling the $4\times 4$ BAS dataset on which it was trained (not shown). In this first study, we have decided to omit any regularization of the learning process. We have done this to keep the focus as clear as possible on the potential gains obtained by using QuALe and to avoid the search of optimal regularization parameters that could be very expensive due to the time to access the DW2X. While this may lead to drops in likelihood \cite{Fischer-2010}, we expect that the substantial reduction in the number of parameters mentioned above may act as an implicit regularizing sparsity constraint. Since we have neglected regularization altogether in all the learning algorithms, we expect the comparison to be fair. Moreover, as the work by Dumoulin {\it et al.} \cite{Dumolin-2014} suggests, the Chimera-RBM model we have investigated has a limited expressive power. So we have decided to delay the investigation of the role of regularization for when we deal with more expressive models that can be naturally represented in a Chimera topology. 

RBMs have the nice feature that sampling in one layer conditioned to a configuration in the other layer can be done in parallel and in one step; this is one of the main reasons for their wide adoption. This feature does not hold true anymore once we have non-trivial lateral connections in one of the layers, which is the concept behind more powerful Boltzmann machines~\cite{schulz2011exploiting,salakhutdinov2008learning}. We think this is one of the most promising directions to explore with the quantum-assisted learning (QuALe) algorithm. By restricting QuALe to study RBM or Chimera-RBM models, we are paying the price of using a device that is in principle more powerful, but we are not taking advantage of having a more general model. It is important to investigate how to take full advantage of the DW2X by designing more suitable models based on the Chimera graph. An interesting possibility is the one explored in Ref.~\cite{Denil-2011} where the Chimera graph of the DW2X is used as a hidden layer to build a semi-restricted Boltzmann machine, which therefore has lateral connections in the hidden layer. When dealing with more general Boltzmann machines it would be interesting to compare the performance of QuALe against mean field methods. Recently, there has been interest in applying mean field techniques for learning restricted Boltzmann machines too \cite{Huang-PRE-2015, Gabrie-arXiv-2015}. Future work should explore how the performance of mean field techniques compares with QuALe's.
 
However, the goal of this first QuALe implementation on small Chimera-RBMs serves several purposes. When dealing with large datasets the log-likelihood cannot be exhaustively computed due to the intractability of computing the partition function. Log-likelihood is the gold standard metric, but it becomes intractable for large systems. In these cases, other performance metrics such as reconstruction error or cross-entropy error turn out to be more convenient, but although widely used, they are rough approximations to the log-likelihood \cite{Hinton-TechRep-2012}. If we were to use these proxies we could not be sure that we would be drawing the right conclusions. This justifies why we used a moderately small dataset with 16 visible and 16 hidden units, and even though computing the log-likelihood was computationally expensive for the study performed here, having 32 units in total was still a manageable size. Through the computation of the exact likelihood we were able to examine in more detail some of the goals proposed here: being able to assess the best effective temperature fit to the desired Boltzmann distribution and to show that using a constant temperature different from the one estimated with our approach might lead to severe suboptimal performance.

Another aspect we explored in this study was to go beyond the conventional CD-1, with the purpose of having a fairer comparison to the results that might be expected from the entirely classical algorithm counterpart. Previous results from our research group \cite{Marc}, as well as others reported by other researchers~\cite{Adachi-arXiv-2015,firstdbm}, are limited to comparing the performance of quantum annealers to the quick but suboptimal {CD-1}. As shown in those studies, even with a suboptimal constant temperature one might be drawn to conclude that QuALe is outperforming conventional CD. Similar conclusions might be drawn from the curves for constant but suboptimal $T=0.08$ and $T_{\rm{av}}\approx 0.1$ vs. CD-1 in Fig.~\ref{f:train}. As shown in Fig.~\ref{f:train}, this conclusion does not hold anymore for higher values of $k$, while the method using the effective-temperature estimation proposed here is the only one showing a steady increase in performance, close to matching the largest value of $k$ tried here, i.e. $k=100$. 

Another important point to investigate in the future is whether the differences observed in performance remain for larger and more complex datasets. We would expect that the performance of CD-$k$ degrades with larger instances as equilibration times are expected to grow fast with the number of variables once the probability distribution starts having non-trivial structure. From this perspective, it is important to notice that QuALe is expected to display a more uniform exploration of configuration space. 

A related important question has to do with the scalability of our temperature estimation technique, i.e. how should the number of samples grow with problem size? In principle, as long as the quantum annealer converges to an approximately Boltzmann distribution and the effective temperature remains constant under rescalings of its control parameters, our method is exact given enough samples. We have left this question for future work as we consider that there are more pressing issues, i.e. limited connectivity and noise, that need to be addressed before we can say something conclusive about scalability. Needless to say, the validity of the assumptions on which our work relies should also be investigated in more detail. It is also important to devise more controlled experiments that allow us to isolate the different phenomena involved. Two months after submission of this manuscript, we learned of ongoing work addressing some of these issues and putting forward other temperature estimation techniques~\cite{Raymond-DWave-2016}. Finally, an investigation on the bias and variance of our effective temperature estimator is an interesting theoretical question that we expect to address in future work.

There are other ways in which the ideas explored here could be extended. For instance, we can go beyond restricted Boltzmann machines to build deep learning architectures or beyond unsupervised learning to build discriminative models. In principle the speed of learning could be increased by adding a `momentum' term to the gradient-ascent learning rule \cite{fischer2012introduction}. Indeed, Adachi and Henderson have started exploring these ideas in a recent work \cite{Adachi-arXiv-2015}. Instead, we have focused on first trying to better understand the basics before adding more (classical) complexity to the learning algorithms that we feel have the risk to obscure the actual contributions from our approach.

\appendix
\section{Comparison to alternative temperature estimation techniques}\label{s:PLE}
Here we discuss alternative techniques to approximately estimate the instance-dependent effective temperature $T_{\rm eff}$, which are in principle efficient too, and show evidence that our method produces superior results. 

One of the mainstream approaches in statistical physics to estimate parameters of an Ising model goes under the name of {\it inverse Ising model}~\cite{Schneidman-Nature-2006,Mezard-Mora-2009,Ricci-Tersenghi-JSTAT-2012,Chau-JSTAT-2012,Aurell-PRL-2012,Decelle-arXiv-2015}. One of the most investigated techniques for solving the inverse Ising model relies on mean field approximations~\cite{Mezard-Mora-2009,Ricci-Tersenghi-JSTAT-2012,Chau-JSTAT-2012}, due to its relative simplicity. These techniques fail, though, for low temperatures where low-energy configurations are arranged in a non-trivial clustered phase~\cite{Decelle-arXiv-2015}. On the other hand, the so-called pseudo-likelihood method~\cite{Aurell-PRL-2012} is recognized as the state-of-the art in solving this problem. Recently, it has been suggested that by suitably introducing information about the clustered phase into mean field methods, these can yield comparable results to the pseudo-likelihood method~\cite{Decelle-arXiv-2015}.  

We first devised a simple strategy to test the feasibility of a mean field approach before attempting to develop a technique specifically targeted to the estimation of $T_{\rm eff}$ alone. Indeed, since we know the control parameters $J_{ij}$ and $h_i$, we can in principle estimate $T_{\rm eff}$ by first determining $W_{ij}$ and $b_i$ using the Bethe approximation~\cite{Ricci-Tersenghi-JSTAT-2012}, and then finding the value of $T_{\rm eff}$ that minimizes some distance between the control parameters and the estimated ones. However, the estimation of $W_{ij}$ and $b_i$ using the samples from the DW2X along the learning path only produces real values up to about the first hundred iterations (not shown). This suggests the Bethe approximation is not suitable for the parameter regime traversed when learning the BAS dataset studied here. 

Since, as we mentioned above, the pseudo-likelihood method ~\cite{Aurell-PRL-2012} is considered the state of the art technique for estimating the parameters of an Ising model we will focus from now on in such an approach. We will see that our method displays a much better performance on the BAS dataset studied here.

Given a set of samples $\mcD =\{\bs^1,\dotsc ,\bs^D\}$, where ${\bs^d = (s_1^d ,\dotsc , s_N^d)}$ with $d=1,\dotsc , D$, generated by a quantum annealer with control parameters $J_{ij}$ and $h_i$, we can estimate the effective temperature $T_{\rm eff}$ by maximizing the {\it average pseudo-likelihood}~\cite{Aurell-PRL-2012}
\begin{widetext}
\begin{equation}\label{e:PLE}
\Lambda(T_{\rm eff}) = -\frac{1}{N\,D}\sum_{i=1}^N\sum_{d=1}^D\ln\left\{1+\exp\left[-\frac{2\, s_i^d}{T_{\rm eff}}\, \left(h_i+\sum_{j\in\partial i}J_{ij}\, s_j^d\right)\right]\right\},
\end{equation}
\end{widetext}
where $\partial i$ denotes the set of neighbors of $i$.

In contrast to the approach in Ref.~\cite{Aurell-PRL-2012}, here the only unknown is $T_{\rm eff}$. We can find a maximum average pseudo-likelihood estimator for the effective temperature ${T_{\rm eff}^{\rm PL} =\arg\max_{T_{\rm eff}} \Lambda(T_{\rm eff})}$ via second order Newton's method~\cite{Aurell-PRL-2012}. In our experiments, we start from $T_{\rm eff}=1$ and iterate until the update is smaller than a tolerance level of $10^{-5}$. Fig.~\ref{f:PL}a shows a comparison of the performance of our quantum-assisted learning algorithm QuALe@$T_{\rm eff}$ with $T_{\rm eff}$ estimated with the pseudo-likelihood method (pink circles) as described here and estimated with the method introduced in Sec.~\ref{s:T} (blue crosses) of the present work. 
We can observe that while QuALe@$T_{\rm eff}$ with the pseudo-likelihood method performs better on the first about 1000 iterations, QuALe@$T_{\rm eff}$ with linear regression performs better afterwards, reaching higher values for the likelihood function. Fig.~\ref{f:PL}b shows the values of effective temperatures estimated by the two techniques along the learning path; interestingly, the effective temperatures estimated by the pseudo-likelihood (pink points on the bottom) are consistently smaller and have less variability than those estimated with our linear regression technique (blue points on the top).

\begin{widetext}

\begin{figure}
\includegraphics[width=0.5\columnwidth]{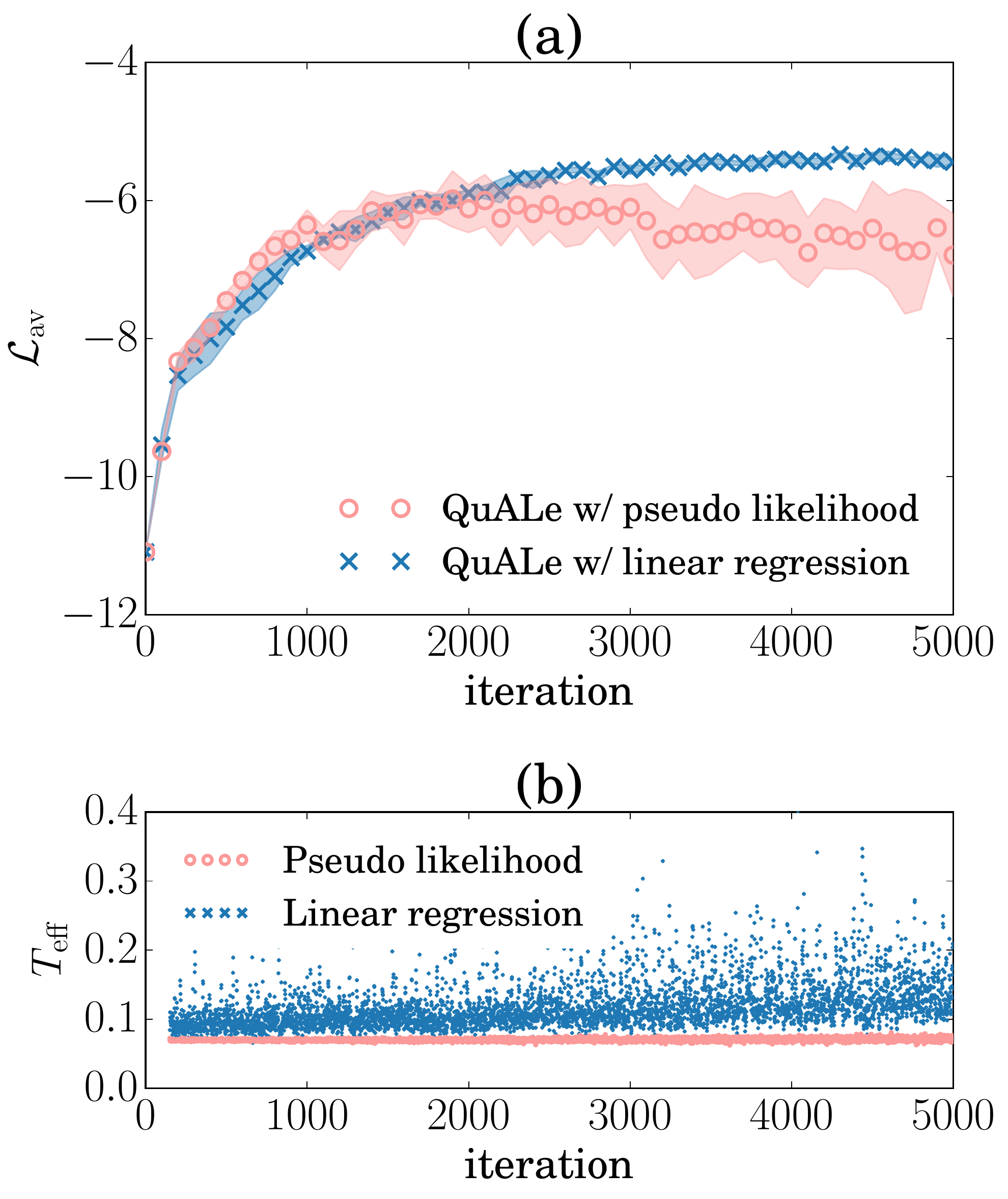}
\caption{{\it  Comparison with pseudo-likelihood estimation:} (a) Performance of the quantum-assisted learning of a Chimera-RBM on the $4\times 4$ BAS dataset using the two different temperature estimation techniques described in Sec.~\ref{s:T}: one (blue crosses) based on linear regression (Eq.~\eqref{e:dl}) and the other (pink circles) on pseudo-likelihood maximization (Eq.~\eqref{e:PLE}). The performance is measured in terms of the average log-likelihood $\mathcal{L}_{\rm av}$, which has been evaluated exhaustively every 50 iterations. The points correspond to the average of $\mathcal{L}_{\rm av}$ over five runs and the bands to one standard deviation.  (b) Variation of the effective temperatures during one of the five runs using linear regression (blue crosses) and pseudo-likelihood maximization (pink circles). Temperature estimation begins at iteration $100$ after restarting from CD-$1$.
}\label{f:PL}
\end{figure}

\section*{Acknowledgments}

This work was supported by NASA Ames Research Center. The authors would like to thank V. M. Janakiraman, Z. Jiang, T. Lanting, E. Rieffel, N. Wiebe, and B. Jacobs for useful discussions. 

\end{widetext}


\begin{thebibliography}{69}
\expandafter\ifx\csname natexlab\endcsname\relax\def\natexlab#1{#1}\fi
\expandafter\ifx\csname bibnamefont\endcsname\relax
  \def\bibnamefont#1{#1}\fi
\expandafter\ifx\csname bibfnamefont\endcsname\relax
  \def\bibfnamefont#1{#1}\fi
\expandafter\ifx\csname citenamefont\endcsname\relax
  \def\citenamefont#1{#1}\fi
\expandafter\ifx\csname url\endcsname\relax
  \def\url#1{\texttt{#1}}\fi
\expandafter\ifx\csname urlprefix\endcsname\relax\def\urlprefix{URL }\fi
\providecommand{\bibinfo}[2]{#2}
\providecommand{\eprint}[2][]{\url{#2}}

\bibitem[{\citenamefont{Neven et~al.}(2008)\citenamefont{Neven, Denchev, Rose,
  and Macready}}]{neven2008training}
\bibinfo{author}{\bibfnamefont{H.}~\bibnamefont{Neven}},
  \bibinfo{author}{\bibfnamefont{V.~S.} \bibnamefont{Denchev}},
  \bibinfo{author}{\bibfnamefont{G.}~\bibnamefont{Rose}}, \bibnamefont{and}
  \bibinfo{author}{\bibfnamefont{W.~G.} \bibnamefont{Macready}},
  \bibinfo{journal}{arXiv:0811.0416}  (\bibinfo{year}{2008}).

\bibitem[{\citenamefont{Neven et~al.}(2009{\natexlab{a}})\citenamefont{Neven,
  Denchev, Drew-Brook, Zhang, Macready, and Rose}}]{neven2009nips}
\bibinfo{author}{\bibfnamefont{H.}~\bibnamefont{Neven}},
  \bibinfo{author}{\bibfnamefont{V.~S.} \bibnamefont{Denchev}},
  \bibinfo{author}{\bibfnamefont{M.}~\bibnamefont{Drew-Brook}},
  \bibinfo{author}{\bibfnamefont{J.}~\bibnamefont{Zhang}},
  \bibinfo{author}{\bibfnamefont{W.~G.} \bibnamefont{Macready}},
  \bibnamefont{and} \bibinfo{author}{\bibfnamefont{G.}~\bibnamefont{Rose}}, in
  \emph{\bibinfo{booktitle}{{Demonstrations at NIPS-09, 24th Annual Conference
  on Neural Information Processing Systems}}}
  (\bibinfo{year}{2009}{\natexlab{a}}), pp. \bibinfo{pages}{1--17}.

\bibitem[{\citenamefont{Neven et~al.}(2009{\natexlab{b}})\citenamefont{Neven,
  Denchev, Rose, and Macready}}]{neven2009training}
\bibinfo{author}{\bibfnamefont{H.}~\bibnamefont{Neven}},
  \bibinfo{author}{\bibfnamefont{V.~S.} \bibnamefont{Denchev}},
  \bibinfo{author}{\bibfnamefont{G.}~\bibnamefont{Rose}}, \bibnamefont{and}
  \bibinfo{author}{\bibfnamefont{W.~G.} \bibnamefont{Macready}},
  \bibinfo{journal}{arXiv:0912.0779}  (\bibinfo{year}{2009}{\natexlab{b}}).

\bibitem[{\citenamefont{Bian et~al.}(2010)\citenamefont{Bian, Chudak, Macready,
  and Rose}}]{bian2010ising}
\bibinfo{author}{\bibfnamefont{Z.}~\bibnamefont{Bian}},
  \bibinfo{author}{\bibfnamefont{F.}~\bibnamefont{Chudak}},
  \bibinfo{author}{\bibfnamefont{W.~G.} \bibnamefont{Macready}},
  \bibnamefont{and} \bibinfo{author}{\bibfnamefont{G.}~\bibnamefont{Rose}},
  \bibinfo{type}{Tech. Rep.}, \bibinfo{institution}{D-Wave Systems}
  (\bibinfo{year}{2010}).

\bibitem[{\citenamefont{Denil and De~Freitas}(2011)}]{Denil-2011}
\bibinfo{author}{\bibfnamefont{M.}~\bibnamefont{Denil}} \bibnamefont{and}
  \bibinfo{author}{\bibfnamefont{N.}~\bibnamefont{De~Freitas}},
  \bibinfo{journal}{NIPS Deep Learning and Unsupervised Feature Learning
  Workshop}  (\bibinfo{year}{2011}).

\bibitem[{\citenamefont{Denchev et~al.}(2012)\citenamefont{Denchev, Ding,
  Vishwanathan, and Neven}}]{denchev2012robust}
\bibinfo{author}{\bibfnamefont{V.~S.} \bibnamefont{Denchev}},
  \bibinfo{author}{\bibfnamefont{N.}~\bibnamefont{Ding}},
  \bibinfo{author}{\bibfnamefont{S.}~\bibnamefont{Vishwanathan}},
  \bibnamefont{and} \bibinfo{author}{\bibfnamefont{H.}~\bibnamefont{Neven}},
  \bibinfo{journal}{arXiv:1205.1148}  (\bibinfo{year}{2012}).

\bibitem[{\citenamefont{Lloyd et~al.}(2013)\citenamefont{Lloyd, Mohseni, and
  Rebentrost}}]{Lloyd-arXiv-2013}
\bibinfo{author}{\bibfnamefont{S.}~\bibnamefont{Lloyd}},
  \bibinfo{author}{\bibfnamefont{M.}~\bibnamefont{Mohseni}}, \bibnamefont{and}
  \bibinfo{author}{\bibfnamefont{P.}~\bibnamefont{Rebentrost}},
  \bibinfo{journal}{arXiv:1307.0411}  (\bibinfo{year}{2013}).

\bibitem[{\citenamefont{Pudenz and Lidar}(2013)}]{Pudenz-QIP-2013}
\bibinfo{author}{\bibfnamefont{K.}~\bibnamefont{Pudenz}} \bibnamefont{and}
  \bibinfo{author}{\bibfnamefont{D.}~\bibnamefont{Lidar}},
  \bibinfo{journal}{Quantum Information Processing}
  \textbf{\bibinfo{volume}{12}}, \bibinfo{pages}{2027} (\bibinfo{year}{2013}).

\bibitem[{\citenamefont{Dumoulin et~al.}(2014)\citenamefont{Dumoulin,
  Goodfellow, Courville, and Bengio}}]{Dumolin-2014}
\bibinfo{author}{\bibfnamefont{V.}~\bibnamefont{Dumoulin}},
  \bibinfo{author}{\bibfnamefont{I.~J.} \bibnamefont{Goodfellow}},
  \bibinfo{author}{\bibfnamefont{A.~C.} \bibnamefont{Courville}},
  \bibnamefont{and} \bibinfo{author}{\bibfnamefont{Y.}~\bibnamefont{Bengio}},
  in \emph{\bibinfo{booktitle}{Proceedings of the Twenty-Eighth {AAAI}
  Conference on Artificial Intelligence, July 27 -31, 2014, Qu{\'{e}}bec City,
  Qu{\'{e}}bec, Canada.}} (\bibinfo{year}{2014}), pp.
  \bibinfo{pages}{1199--1205}.

\bibitem[{\citenamefont{Lloyd et~al.}(2014)\citenamefont{Lloyd, Mohseni, and
  Rebentrost}}]{Lloyd-NatPhys-2014}
\bibinfo{author}{\bibfnamefont{S.}~\bibnamefont{Lloyd}},
  \bibinfo{author}{\bibfnamefont{M.}~\bibnamefont{Mohseni}}, \bibnamefont{and}
  \bibinfo{author}{\bibfnamefont{P.}~\bibnamefont{Rebentrost}},
  \bibinfo{journal}{Nature Physics}  (\bibinfo{year}{2014}).

\bibitem[{\citenamefont{Rebentrost et~al.}(2014)\citenamefont{Rebentrost,
  Mohseni, and Lloyd}}]{Rebentrost-PRL-2014}
\bibinfo{author}{\bibfnamefont{P.}~\bibnamefont{Rebentrost}},
  \bibinfo{author}{\bibfnamefont{M.}~\bibnamefont{Mohseni}}, \bibnamefont{and}
  \bibinfo{author}{\bibfnamefont{S.}~\bibnamefont{Lloyd}},
  \bibinfo{journal}{Phys. Rev. Lett.} \textbf{\bibinfo{volume}{113}},
  \bibinfo{pages}{130503} (\bibinfo{year}{2014}).

\bibitem[{\citenamefont{Nathan~Wiebe}(2015)}]{Wiebe-arXiv-2015}
\bibinfo{author}{\bibfnamefont{K.~M.~S.} \bibnamefont{Nathan~Wiebe},
  \bibfnamefont{Ashish~Kapoor}}, \bibinfo{journal}{arXiv:1412.3489}
  (\bibinfo{year}{2015}).

\bibitem[{\citenamefont{Aaronson}(2015)}]{Aaronson-2015}
\bibinfo{author}{\bibfnamefont{S.}~\bibnamefont{Aaronson}},
  \bibinfo{journal}{Nature Physics} \textbf{\bibinfo{volume}{11}},
  \bibinfo{pages}{291} (\bibinfo{year}{2015}), \bibinfo{note}{commentary}.

\bibitem[{\citenamefont{Adachi and Henderson}(2015)}]{Adachi-arXiv-2015}
\bibinfo{author}{\bibfnamefont{S.~H.} \bibnamefont{Adachi}} \bibnamefont{and}
  \bibinfo{author}{\bibfnamefont{M.~P.} \bibnamefont{Henderson}},
  \bibinfo{journal}{arXiv:1510.06356}  (\bibinfo{year}{2015}).

\bibitem[{\citenamefont{Chancellor et~al.}(2016)\citenamefont{Chancellor,
  Szoke, Vinci, Aeppli, and Warburton}}]{chancellor2016maximum}
\bibinfo{author}{\bibfnamefont{N.}~\bibnamefont{Chancellor}},
  \bibinfo{author}{\bibfnamefont{S.}~\bibnamefont{Szoke}},
  \bibinfo{author}{\bibfnamefont{W.}~\bibnamefont{Vinci}},
  \bibinfo{author}{\bibfnamefont{G.}~\bibnamefont{Aeppli}}, \bibnamefont{and}
  \bibinfo{author}{\bibfnamefont{P.~A.} \bibnamefont{Warburton}},
  \bibinfo{journal}{Scientific Reports} \textbf{\bibinfo{volume}{6}}
  (\bibinfo{year}{2016}).

\bibitem[{\citenamefont{{Mohammad H. Amin and Evgeny Andriyash and Jason Rolfe
  and Bohdan Kulchytskyy and Roger Melko}}(2016)}]{Amin-arXiv-2016}
\bibinfo{author}{\bibnamefont{{Mohammad H. Amin and Evgeny Andriyash and Jason
  Rolfe and Bohdan Kulchytskyy and Roger Melko}}},
  \bibinfo{journal}{arXiv:1601.02036}  (\bibinfo{year}{2016}).

\bibitem[{\citenamefont{Finnila et~al.}(1994)\citenamefont{Finnila, Gomez,
  Sebenik, Stenson, and Doll}}]{finnila1994quantum}
\bibinfo{author}{\bibfnamefont{A.}~\bibnamefont{Finnila}},
  \bibinfo{author}{\bibfnamefont{M.}~\bibnamefont{Gomez}},
  \bibinfo{author}{\bibfnamefont{C.}~\bibnamefont{Sebenik}},
  \bibinfo{author}{\bibfnamefont{C.}~\bibnamefont{Stenson}}, \bibnamefont{and}
  \bibinfo{author}{\bibfnamefont{J.}~\bibnamefont{Doll}},
  \bibinfo{journal}{{Chemical Physics Letters}} \textbf{\bibinfo{volume}{219}},
  \bibinfo{pages}{343} (\bibinfo{year}{1994}).

\bibitem[{\citenamefont{Kadowaki and Nishimori}(1998)}]{kadowaki_quantum_1998}
\bibinfo{author}{\bibfnamefont{T.}~\bibnamefont{Kadowaki}} \bibnamefont{and}
  \bibinfo{author}{\bibfnamefont{H.}~\bibnamefont{Nishimori}},
  \bibinfo{journal}{Phys. Rev. E.} \textbf{\bibinfo{volume}{58}},
  \bibinfo{pages}{5355} (\bibinfo{year}{1998}).

\bibitem[{\citenamefont{Farhi et~al.}(2001)\citenamefont{Farhi, Goldstone,
  Gutmann, Lapan, Lundgren, and Preda}}]{Farhi2001}
\bibinfo{author}{\bibfnamefont{E.}~\bibnamefont{Farhi}},
  \bibinfo{author}{\bibfnamefont{J.}~\bibnamefont{Goldstone}},
  \bibinfo{author}{\bibfnamefont{S.}~\bibnamefont{Gutmann}},
  \bibinfo{author}{\bibfnamefont{J.}~\bibnamefont{Lapan}},
  \bibinfo{author}{\bibfnamefont{A.}~\bibnamefont{Lundgren}}, \bibnamefont{and}
  \bibinfo{author}{\bibfnamefont{D.}~\bibnamefont{Preda}},
  \bibinfo{journal}{Science} \textbf{\bibinfo{volume}{292}},
  \bibinfo{pages}{472} (\bibinfo{year}{2001}).

\bibitem[{\citenamefont{Gaitan and Clark}(2012)}]{Gaitan2012}
\bibinfo{author}{\bibfnamefont{F.}~\bibnamefont{Gaitan}} \bibnamefont{and}
  \bibinfo{author}{\bibfnamefont{L.}~\bibnamefont{Clark}},
  \bibinfo{journal}{Phys. Rev. Lett.} \textbf{\bibinfo{volume}{108}},
  \bibinfo{pages}{010501} (\bibinfo{year}{2012}).

\bibitem[{\citenamefont{Perdomo-Ortiz et~al.}(2012)\citenamefont{Perdomo-Ortiz,
  Dickson, Drew-Brook, Rose, and Aspuru-Guzik}}]{PerdomoOrtiz2012_LPF}
\bibinfo{author}{\bibfnamefont{A.}~\bibnamefont{Perdomo-Ortiz}},
  \bibinfo{author}{\bibfnamefont{N.}~\bibnamefont{Dickson}},
  \bibinfo{author}{\bibfnamefont{M.}~\bibnamefont{Drew-Brook}},
  \bibinfo{author}{\bibfnamefont{G.}~\bibnamefont{Rose}}, \bibnamefont{and}
  \bibinfo{author}{\bibfnamefont{A.}~\bibnamefont{Aspuru-Guzik}},
  \bibinfo{journal}{Sci. Rep.} \textbf{\bibinfo{volume}{2}},
  \bibinfo{pages}{571} (\bibinfo{year}{2012}).

\bibitem[{\citenamefont{Bian et~al.}(2014)\citenamefont{Bian, Chudak, Israel,
  Lackey, Macready, and Roy}}]{Bian2014}
\bibinfo{author}{\bibfnamefont{Z.}~\bibnamefont{Bian}},
  \bibinfo{author}{\bibfnamefont{F.}~\bibnamefont{Chudak}},
  \bibinfo{author}{\bibfnamefont{R.}~\bibnamefont{Israel}},
  \bibinfo{author}{\bibfnamefont{B.}~\bibnamefont{Lackey}},
  \bibinfo{author}{\bibfnamefont{W.~G.} \bibnamefont{Macready}},
  \bibnamefont{and} \bibinfo{author}{\bibfnamefont{A.}~\bibnamefont{Roy}},
  \bibinfo{journal}{Frontiers in Physics} \textbf{\bibinfo{volume}{2}}
  (\bibinfo{year}{2014}), ISSN \bibinfo{issn}{2296-424X}.

\bibitem[{\citenamefont{O'Gorman et~al.}(2015)\citenamefont{O'Gorman, Babbush,
  Perdomo-Ortiz, Aspuru-Guzik, and Smelyanskiy}}]{OGorman-EPJST-2015}
\bibinfo{author}{\bibfnamefont{B.}~\bibnamefont{O'Gorman}},
  \bibinfo{author}{\bibfnamefont{R.}~\bibnamefont{Babbush}},
  \bibinfo{author}{\bibfnamefont{A.}~\bibnamefont{Perdomo-Ortiz}},
  \bibinfo{author}{\bibfnamefont{A.}~\bibnamefont{Aspuru-Guzik}},
  \bibnamefont{and}
  \bibinfo{author}{\bibfnamefont{V.}~\bibnamefont{Smelyanskiy}},
  \bibinfo{journal}{The European Physical Journal Special Topics}
  \textbf{\bibinfo{volume}{224}}, \bibinfo{pages}{163} (\bibinfo{year}{2015}),
  ISSN \bibinfo{issn}{1951-6355}.

\bibitem[{\citenamefont{Rieffel et~al.}(2015)\citenamefont{Rieffel, Venturelli,
  O'Gorman, Do, Prystay, and Smelyanskiy}}]{RieffelQIP2015}
\bibinfo{author}{\bibfnamefont{E.~G.} \bibnamefont{Rieffel}},
  \bibinfo{author}{\bibfnamefont{D.}~\bibnamefont{Venturelli}},
  \bibinfo{author}{\bibfnamefont{B.}~\bibnamefont{O'Gorman}},
  \bibinfo{author}{\bibfnamefont{M.~B.} \bibnamefont{Do}},
  \bibinfo{author}{\bibfnamefont{E.~M.} \bibnamefont{Prystay}},
  \bibnamefont{and} \bibinfo{author}{\bibfnamefont{V.~N.}
  \bibnamefont{Smelyanskiy}}, \bibinfo{journal}{Quantum Information Processing}
  \textbf{\bibinfo{volume}{14}}, \bibinfo{pages}{1} (\bibinfo{year}{2015}),
  ISSN \bibinfo{issn}{1570-0755}.

\bibitem[{\citenamefont{{Perdomo-Ortiz, A.}
  et~al.}(2015)\citenamefont{{Perdomo-Ortiz, A.}, {Fluegemann, J.},
  {Narasimhan, S.}, {Biswas, R.}, and {Smelyanskiy,
  V.N.}}}]{PerdomoOrtiz_EPJST2015}
\bibinfo{author}{\bibnamefont{{Perdomo-Ortiz, A.}}},
  \bibinfo{author}{\bibnamefont{{Fluegemann, J.}}},
  \bibinfo{author}{\bibnamefont{{Narasimhan, S.}}},
  \bibinfo{author}{\bibnamefont{{Biswas, R.}}}, \bibnamefont{and}
  \bibinfo{author}{\bibnamefont{{Smelyanskiy, V.N.}}}, \bibinfo{journal}{Eur.
  Phys. J. Special Topics} \textbf{\bibinfo{volume}{224}}, \bibinfo{pages}{131}
  (\bibinfo{year}{2015}).

\bibitem[{\citenamefont{Perdomo-Ortiz et~al.}(2015)\citenamefont{Perdomo-Ortiz,
  Fluegemann, Biswas, and Smelyanskiy}}]{perdomo2015performance}
\bibinfo{author}{\bibfnamefont{A.}~\bibnamefont{Perdomo-Ortiz}},
  \bibinfo{author}{\bibfnamefont{J.}~\bibnamefont{Fluegemann}},
  \bibinfo{author}{\bibfnamefont{R.}~\bibnamefont{Biswas}}, \bibnamefont{and}
  \bibinfo{author}{\bibfnamefont{V.~N.} \bibnamefont{Smelyanskiy}},
  \bibinfo{journal}{arXiv:1503.01083}  (\bibinfo{year}{2015}).

\bibitem[{\citenamefont{Venturelli et~al.}(2015)\citenamefont{Venturelli,
  Marchand, and Rojo}}]{Venturelli-JobShop-arXiv-2015}
\bibinfo{author}{\bibfnamefont{D.}~\bibnamefont{Venturelli}},
  \bibinfo{author}{\bibfnamefont{D.~J.} \bibnamefont{Marchand}},
  \bibnamefont{and} \bibinfo{author}{\bibfnamefont{G.}~\bibnamefont{Rojo}},
  \bibinfo{journal}{arXiv:1506.08479}  (\bibinfo{year}{2015}).

\bibitem[{\citenamefont{Amin}(2015)}]{Amin-arXiv-2015}
\bibinfo{author}{\bibfnamefont{M.~H.} \bibnamefont{Amin}},
  \bibinfo{journal}{Phys. Rev. A} \textbf{\bibinfo{volume}{92}},
  \bibinfo{pages}{052323} (\bibinfo{year}{2015}).

\bibitem[{\citenamefont{Perdomo-Ortiz et~al.}(2016)\citenamefont{Perdomo-Ortiz,
  O'Gorman, Fluegemann, Biswas, and Smelyanskiy}}]{PerdomoOrtiz_SciRep2016}
\bibinfo{author}{\bibfnamefont{A.}~\bibnamefont{Perdomo-Ortiz}},
  \bibinfo{author}{\bibfnamefont{B.}~\bibnamefont{O'Gorman}},
  \bibinfo{author}{\bibfnamefont{J.}~\bibnamefont{Fluegemann}},
  \bibinfo{author}{\bibfnamefont{R.}~\bibnamefont{Biswas}}, \bibnamefont{and}
  \bibinfo{author}{\bibfnamefont{V.~N.} \bibnamefont{Smelyanskiy}},
  \bibinfo{journal}{Sci. Rep.} \textbf{\bibinfo{volume}{6}},
  \bibinfo{pages}{18628} (\bibinfo{year}{2016}).

\bibitem[{\citenamefont{Dorband}(2015)}]{Dorband}
\bibinfo{author}{\bibfnamefont{J.~E.} \bibnamefont{Dorband}}, in
  \emph{\bibinfo{booktitle}{ITNG, Washington DC}} (\bibinfo{year}{2015}), pp.
  \bibinfo{pages}{703--707}.

\bibitem[{\citenamefont{Sinclair and Jerrum}(1989)}]{Sinclair-InfComp-1989}
\bibinfo{author}{\bibfnamefont{A.}~\bibnamefont{Sinclair}} \bibnamefont{and}
  \bibinfo{author}{\bibfnamefont{M.}~\bibnamefont{Jerrum}},
  \bibinfo{journal}{Inf. Comput.} \textbf{\bibinfo{volume}{82}},
  \bibinfo{pages}{93} (\bibinfo{year}{1989}), ISSN \bibinfo{issn}{0890-5401},
  \urlprefix\url{http://dx.doi.org/10.1016/0890-5401(89)90067-9}.

\bibitem[{\citenamefont{Frigessi et~al.}(1997)\citenamefont{Frigessi,
  Martinelli, and Stander}}]{Frigessi-Biometrika-1997}
\bibinfo{author}{\bibfnamefont{A.}~\bibnamefont{Frigessi}},
  \bibinfo{author}{\bibfnamefont{F.}~\bibnamefont{Martinelli}},
  \bibnamefont{and} \bibinfo{author}{\bibfnamefont{J.}~\bibnamefont{Stander}},
  \bibinfo{journal}{Biometrika} \textbf{\bibinfo{volume}{84}},
  \bibinfo{pages}{1} (\bibinfo{year}{1997}).

\bibitem[{\citenamefont{Long and Servedio}(2010)}]{Long-2010}
\bibinfo{author}{\bibfnamefont{P.~M.} \bibnamefont{Long}} \bibnamefont{and}
  \bibinfo{author}{\bibfnamefont{R.}~\bibnamefont{Servedio}}, in
  \emph{\bibinfo{booktitle}{Proceedings of the 27th International Conference on
  Machine Learning (ICML-10)}}, edited by
  \bibinfo{editor}{\bibfnamefont{J.}~\bibnamefont{Fürnkranz}}
  \bibnamefont{and} \bibinfo{editor}{\bibfnamefont{T.}~\bibnamefont{Joachims}}
  (\bibinfo{publisher}{Omnipress}, \bibinfo{year}{2010}), pp.
  \bibinfo{pages}{703--710}.

\bibitem[{\citenamefont{LeCun et~al.}(2015)\citenamefont{LeCun, Bengio, and
  Hinton}}]{LeCun-Nature-2015}
\bibinfo{author}{\bibfnamefont{Y.}~\bibnamefont{LeCun}},
  \bibinfo{author}{\bibfnamefont{Y.}~\bibnamefont{Bengio}}, \bibnamefont{and}
  \bibinfo{author}{\bibfnamefont{G.}~\bibnamefont{Hinton}},
  \bibinfo{journal}{Nature} \textbf{\bibinfo{volume}{521}}, \bibinfo{pages}{436
  } (\bibinfo{year}{2015}).

\bibitem[{\citenamefont{Smolensky}(1986)}]{Smolensky-RBM-1986}
\bibinfo{author}{\bibfnamefont{P.}~\bibnamefont{Smolensky}}, in
  \emph{\bibinfo{booktitle}{Parallel Distributed Processing: Explorations in
  the Microstructure of Cognition, Vol. 1}}, edited by
  \bibinfo{editor}{\bibfnamefont{D.~E.} \bibnamefont{Rumelhart}},
  \bibinfo{editor}{\bibfnamefont{J.~L.} \bibnamefont{McClelland}},
  \bibnamefont{and} \bibinfo{editor}{\bibfnamefont{C.}~\bibnamefont{PDP
  Research~Group}} (\bibinfo{publisher}{MIT Press},
  \bibinfo{address}{Cambridge, MA, USA}, \bibinfo{year}{1986}), chap.
  \bibinfo{chapter}{Information Processing in Dynamical Systems: Foundations of
  Harmony Theory}, pp. \bibinfo{pages}{194--281}, ISBN
  \bibinfo{isbn}{0-262-68053-X}.

\bibitem[{\citenamefont{Hinton}(2002)}]{hinton2002training}
\bibinfo{author}{\bibfnamefont{G.~E.} \bibnamefont{Hinton}},
  \bibinfo{journal}{{Neural Computation}} \textbf{\bibinfo{volume}{14}},
  \bibinfo{pages}{1771} (\bibinfo{year}{2002}).

\bibitem[{\citenamefont{Hinton and Sejnowski}(1986)}]{Hinton-BAS-1986}
\bibinfo{author}{\bibfnamefont{G.~E.} \bibnamefont{Hinton}} \bibnamefont{and}
  \bibinfo{author}{\bibfnamefont{T.~J.} \bibnamefont{Sejnowski}}
  (\bibinfo{publisher}{MIT Press}, \bibinfo{address}{Cambridge, MA, USA},
  \bibinfo{year}{1986}), chap. \bibinfo{chapter}{Learning and Relearning in
  Boltzmann Machines}, pp. \bibinfo{pages}{282--317}, ISBN
  \bibinfo{isbn}{0-262-68053-X}.

\bibitem[{\citenamefont{MacKay}(2002)}]{MacKay-book-2002}
\bibinfo{author}{\bibfnamefont{D.~J.~C.} \bibnamefont{MacKay}},
  \emph{\bibinfo{title}{Information Theory, Inference \& Learning Algorithms}}
  (\bibinfo{publisher}{Cambridge University Press}, \bibinfo{address}{New York,
  NY, USA}, \bibinfo{year}{2002}), ISBN \bibinfo{isbn}{0521642981}.

\bibitem[{\citenamefont{Fischer and Igel}(2012)}]{fischer2012introduction}
\bibinfo{author}{\bibfnamefont{A.}~\bibnamefont{Fischer}} \bibnamefont{and}
  \bibinfo{author}{\bibfnamefont{C.}~\bibnamefont{Igel}}, in
  \emph{\bibinfo{booktitle}{Progress in Pattern Recognition, Image Analysis,
  Computer Vision, and Applications}} (\bibinfo{publisher}{Springer},
  \bibinfo{year}{2012}), pp. \bibinfo{pages}{14--36}.

\bibitem[{\citenamefont{Le~Roux and Bengio}(2008)}]{LeRoux-2008}
\bibinfo{author}{\bibfnamefont{N.}~\bibnamefont{Le~Roux}} \bibnamefont{and}
  \bibinfo{author}{\bibfnamefont{Y.}~\bibnamefont{Bengio}},
  \bibinfo{journal}{Neural Computation} \textbf{\bibinfo{volume}{20}},
  \bibinfo{pages}{1631} (\bibinfo{year}{2008}), ISSN \bibinfo{issn}{0899-7667}.

\bibitem[{\citenamefont{Johnson et~al.}(2011)\citenamefont{Johnson, Amin,
  Gildert, Lanting, Hamze, Dickson, Harris, Berkley, Johansson, Bunyk
  et~al.}}]{Johnson-Nature-2011}
\bibinfo{author}{\bibfnamefont{M.~W.} \bibnamefont{Johnson}},
  \bibinfo{author}{\bibfnamefont{M.~H.~S.} \bibnamefont{Amin}},
  \bibinfo{author}{\bibfnamefont{S.}~\bibnamefont{Gildert}},
  \bibinfo{author}{\bibfnamefont{T.}~\bibnamefont{Lanting}},
  \bibinfo{author}{\bibfnamefont{F.}~\bibnamefont{Hamze}},
  \bibinfo{author}{\bibfnamefont{N.}~\bibnamefont{Dickson}},
  \bibinfo{author}{\bibfnamefont{R.}~\bibnamefont{Harris}},
  \bibinfo{author}{\bibfnamefont{A.~J.} \bibnamefont{Berkley}},
  \bibinfo{author}{\bibfnamefont{J.}~\bibnamefont{Johansson}},
  \bibinfo{author}{\bibfnamefont{P.}~\bibnamefont{Bunyk}},
  \bibnamefont{et~al.}, \bibinfo{journal}{Nature}
  \textbf{\bibinfo{volume}{473}}, \bibinfo{pages}{194} (\bibinfo{year}{2011}),
  ISSN \bibinfo{issn}{0028-0836}.

\bibitem[{\citenamefont{Harris et~al.}(2010)\citenamefont{Harris, Johnson,
  Lanting, Berkley, Johansson, Bunyk, Tolkacheva, Ladizinsky, Ladizinsky, Oh
  et~al.}}]{Harris-PRB-2010}
\bibinfo{author}{\bibfnamefont{R.}~\bibnamefont{Harris}},
  \bibinfo{author}{\bibfnamefont{M.~W.} \bibnamefont{Johnson}},
  \bibinfo{author}{\bibfnamefont{T.}~\bibnamefont{Lanting}},
  \bibinfo{author}{\bibfnamefont{A.~J.} \bibnamefont{Berkley}},
  \bibinfo{author}{\bibfnamefont{J.}~\bibnamefont{Johansson}},
  \bibinfo{author}{\bibfnamefont{P.}~\bibnamefont{Bunyk}},
  \bibinfo{author}{\bibfnamefont{E.}~\bibnamefont{Tolkacheva}},
  \bibinfo{author}{\bibfnamefont{E.}~\bibnamefont{Ladizinsky}},
  \bibinfo{author}{\bibfnamefont{N.}~\bibnamefont{Ladizinsky}},
  \bibinfo{author}{\bibfnamefont{T.}~\bibnamefont{Oh}}, \bibnamefont{et~al.},
  \bibinfo{journal}{Phys. Rev. B} \textbf{\bibinfo{volume}{82}},
  \bibinfo{pages}{024511} (\bibinfo{year}{2010}).

\bibitem[{\citenamefont{Shor}(1994)}]{Shor1994}
\bibinfo{author}{\bibfnamefont{P.}~\bibnamefont{Shor}}, in
  \emph{\bibinfo{booktitle}{Foundations of Computer Science, 1994 Proceedings.,
  35th Annual Symposium on}} (\bibinfo{year}{1994}), pp.
  \bibinfo{pages}{124--134}.

\bibitem[{\citenamefont{Bengio and Delalleau}(2009)}]{Bengio+Delalleau-2009}
\bibinfo{author}{\bibfnamefont{Y.}~\bibnamefont{Bengio}} \bibnamefont{and}
  \bibinfo{author}{\bibfnamefont{O.}~\bibnamefont{Delalleau}},
  \bibinfo{journal}{Neural Computation} \textbf{\bibinfo{volume}{21}},
  \bibinfo{pages}{1601} (\bibinfo{year}{2009}).

\bibitem[{\citenamefont{Fischer and Igel}(2010)}]{Fischer-2010}
\bibinfo{author}{\bibfnamefont{A.}~\bibnamefont{Fischer}} \bibnamefont{and}
  \bibinfo{author}{\bibfnamefont{C.}~\bibnamefont{Igel}}, in
  \emph{\bibinfo{booktitle}{Proceedings of the 20th International Conference on
  Artificial Neural Networks: Part III}} (\bibinfo{publisher}{Springer-Verlag},
  \bibinfo{address}{Berlin, Heidelberg}, \bibinfo{year}{2010}), ICANN'10, pp.
  \bibinfo{pages}{208--217}, ISBN \bibinfo{isbn}{3-642-15824-2,
  978-3-642-15824-7}.

\bibitem[{\citenamefont{Goodfellow et~al.}(2016)\citenamefont{Goodfellow,
  Bengio, and Courville}}]{Bengio-Book}
\bibinfo{author}{\bibfnamefont{I.}~\bibnamefont{Goodfellow}},
  \bibinfo{author}{\bibfnamefont{Y.}~\bibnamefont{Bengio}}, \bibnamefont{and}
  \bibinfo{author}{\bibfnamefont{A.}~\bibnamefont{Courville}}
  (\bibinfo{year}{2016}), \bibinfo{note}{book in preparation for MIT Press},
  \urlprefix\url{http://goodfeli.github.io/dlbook/}.

\bibitem[{\citenamefont{Albash et~al.}(2012)\citenamefont{Albash, Boixo, Lidar,
  and Zanardi}}]{Albash_NJP2012}
\bibinfo{author}{\bibfnamefont{T.}~\bibnamefont{Albash}},
  \bibinfo{author}{\bibfnamefont{S.}~\bibnamefont{Boixo}},
  \bibinfo{author}{\bibfnamefont{D.~A.} \bibnamefont{Lidar}}, \bibnamefont{and}
  \bibinfo{author}{\bibfnamefont{P.}~\bibnamefont{Zanardi}},
  \bibinfo{journal}{New Journal of Physics} \textbf{\bibinfo{volume}{14}},
  \bibinfo{pages}{123016} (\bibinfo{year}{2012}).

\bibitem[{\citenamefont{Smelyanskiy et~al.}(2015)\citenamefont{Smelyanskiy,
  Venturelli, Perdomo-Ortiz, Knysh, and Dykman}}]{Smelyanskiy_arXiv2015}
\bibinfo{author}{\bibfnamefont{V.~N.} \bibnamefont{Smelyanskiy}},
  \bibinfo{author}{\bibfnamefont{D.}~\bibnamefont{Venturelli}},
  \bibinfo{author}{\bibfnamefont{A.}~\bibnamefont{Perdomo-Ortiz}},
  \bibinfo{author}{\bibfnamefont{S.}~\bibnamefont{Knysh}}, \bibnamefont{and}
  \bibinfo{author}{\bibfnamefont{M.~I.} \bibnamefont{Dykman}},
  \bibinfo{journal}{arXiv:1511.02581}  (\bibinfo{year}{2015}).

\bibitem[{\citenamefont{Boixo et~al.}(2014)\citenamefont{Boixo, Smelyanskiy,
  Shabani, Isakov, Dykman, Denchev, Amin, Smirnov, Mohseni, and
  Neven}}]{boixo2014computational}
\bibinfo{author}{\bibfnamefont{S.}~\bibnamefont{Boixo}},
  \bibinfo{author}{\bibfnamefont{V.~N.} \bibnamefont{Smelyanskiy}},
  \bibinfo{author}{\bibfnamefont{A.}~\bibnamefont{Shabani}},
  \bibinfo{author}{\bibfnamefont{S.~V.} \bibnamefont{Isakov}},
  \bibinfo{author}{\bibfnamefont{M.}~\bibnamefont{Dykman}},
  \bibinfo{author}{\bibfnamefont{V.~S.} \bibnamefont{Denchev}},
  \bibinfo{author}{\bibfnamefont{M.}~\bibnamefont{Amin}},
  \bibinfo{author}{\bibfnamefont{A.}~\bibnamefont{Smirnov}},
  \bibinfo{author}{\bibfnamefont{M.}~\bibnamefont{Mohseni}}, \bibnamefont{and}
  \bibinfo{author}{\bibfnamefont{H.}~\bibnamefont{Neven}},
  \bibinfo{journal}{Nature Communications} \textbf{\bibinfo{volume}{7}},
  \bibinfo{pages}{10327} (\bibinfo{year}{2014}).

\bibitem[{\citenamefont{Balasubramanian}(1997)}]{Balasubramanian-1997}
\bibinfo{author}{\bibfnamefont{V.}~\bibnamefont{Balasubramanian}},
  \bibinfo{journal}{Neural Comput.} \textbf{\bibinfo{volume}{9}},
  \bibinfo{pages}{349} (\bibinfo{year}{1997}), ISSN \bibinfo{issn}{0899-7667}.

\bibitem[{\citenamefont{Myung et~al.}(2000)\citenamefont{Myung,
  Balasubramanian, and Pitt}}]{Myung-PNAS-2000}
\bibinfo{author}{\bibfnamefont{I.~J.} \bibnamefont{Myung}},
  \bibinfo{author}{\bibfnamefont{V.}~\bibnamefont{Balasubramanian}},
  \bibnamefont{and} \bibinfo{author}{\bibfnamefont{M.~A.} \bibnamefont{Pitt}},
  \bibinfo{journal}{Proceedings of the National Academy of Sciences}
  \textbf{\bibinfo{volume}{97}}, \bibinfo{pages}{11170} (\bibinfo{year}{2000}).

\bibitem[{\citenamefont{{Mastromatteo}}(2013)}]{Mastromatteo-PhD-2013}
\bibinfo{author}{\bibfnamefont{I.}~\bibnamefont{{Mastromatteo}}},
  \bibinfo{journal}{ArXiv e-prints}  (\bibinfo{year}{2013}),
  \eprint{1311.0190}.

\bibitem[{\citenamefont{Habeck}(2015)}]{Habeck-arXiv-2015}
\bibinfo{author}{\bibfnamefont{M.}~\bibnamefont{Habeck}},
  \bibinfo{journal}{arXiv:1504.00053}  (\bibinfo{year}{2015}).

\bibitem[{\citenamefont{Bishop}(2006)}]{Bishop-book-2006}
\bibinfo{author}{\bibfnamefont{C.~M.} \bibnamefont{Bishop}},
  \emph{\bibinfo{title}{Pattern Recognition and Machine Learning (Information
  Science and Statistics)}} (\bibinfo{publisher}{Springer-Verlag New York,
  Inc.}, \bibinfo{address}{Secaucus, NJ, USA}, \bibinfo{year}{2006}).

\bibitem[{\citenamefont{Bengio}(2009)}]{Bengio-book-2009}
\bibinfo{author}{\bibfnamefont{Y.}~\bibnamefont{Bengio}},
  \bibinfo{journal}{Found. Trends Mach. Learn.} \textbf{\bibinfo{volume}{2}},
  \bibinfo{pages}{1} (\bibinfo{year}{2009}).

\bibitem[{\citenamefont{Schulz et~al.}(2010)\citenamefont{Schulz, M\"{u}ller,
  and Behnke}}]{Schulz-2010}
\bibinfo{author}{\bibfnamefont{H.}~\bibnamefont{Schulz}},
  \bibinfo{author}{\bibfnamefont{A.~C.} \bibnamefont{M\"{u}ller}},
  \bibnamefont{and} \bibinfo{author}{\bibfnamefont{S.}~\bibnamefont{Behnke}},
  in \emph{\bibinfo{booktitle}{ESANN}} (\bibinfo{year}{2010}).

\bibitem[{\citenamefont{Salakhutdinov}(2008)}]{salakhutdinov2008learning}
\bibinfo{author}{\bibfnamefont{R.}~\bibnamefont{Salakhutdinov}},
  \bibinfo{type}{Tech. Rep.} (\bibinfo{year}{2008}).

\bibitem[{\citenamefont{Huang and Toyoizumi}(2015)}]{Huang-PRE-2015}
\bibinfo{author}{\bibfnamefont{H.}~\bibnamefont{Huang}} \bibnamefont{and}
  \bibinfo{author}{\bibfnamefont{T.}~\bibnamefont{Toyoizumi}},
  \bibinfo{journal}{Phys. Rev. E} \textbf{\bibinfo{volume}{91}},
  \bibinfo{pages}{050101} (\bibinfo{year}{2015}).

\bibitem[{\citenamefont{Gabrie et~al.}(2015)\citenamefont{Gabrie, Tramel, and
  Krzakala}}]{Gabrie-arXiv-2015}
\bibinfo{author}{\bibfnamefont{M.}~\bibnamefont{Gabrie}},
  \bibinfo{author}{\bibfnamefont{E.~W.} \bibnamefont{Tramel}},
  \bibnamefont{and} \bibinfo{author}{\bibfnamefont{F.}~\bibnamefont{Krzakala}},
  in \emph{\bibinfo{booktitle}{Advances in Neural Information Processing
  Systems 28}}, edited by
  \bibinfo{editor}{\bibfnamefont{C.}~\bibnamefont{Cortes}},
  \bibinfo{editor}{\bibfnamefont{N.~D.} \bibnamefont{Lawrence}},
  \bibinfo{editor}{\bibfnamefont{D.~D.} \bibnamefont{Lee}},
  \bibinfo{editor}{\bibfnamefont{M.}~\bibnamefont{Sugiyama}}, \bibnamefont{and}
  \bibinfo{editor}{\bibfnamefont{R.}~\bibnamefont{Garnett}}
  (\bibinfo{publisher}{Curran Associates, Inc.}, \bibinfo{year}{2015}), pp.
  \bibinfo{pages}{640--648}.

\bibitem[{\citenamefont{Hinton}(2012)}]{Hinton-TechRep-2012}
\bibinfo{author}{\bibfnamefont{G.~E.} \bibnamefont{Hinton}}, in
  \emph{\bibinfo{booktitle}{Neural Networks: Tricks of the Trade (2nd ed.)}},
  edited by \bibinfo{editor}{\bibfnamefont{G.}~\bibnamefont{Montavon}},
  \bibinfo{editor}{\bibfnamefont{G.~B.} \bibnamefont{Orr}}, \bibnamefont{and}
  \bibinfo{editor}{\bibfnamefont{K.-R.} \bibnamefont{Müller}}
  (\bibinfo{publisher}{Springer}, \bibinfo{year}{2012}), vol.
  \bibinfo{volume}{7700} of \emph{\bibinfo{series}{Lecture Notes in Computer
  Science}}, pp. \bibinfo{pages}{599--619}.

\bibitem[{\citenamefont{Benedetti}(2015)}]{Marc}
\bibinfo{author}{\bibfnamefont{M.}~\bibnamefont{Benedetti}}, Master's thesis,
  \bibinfo{school}{Universit{\'e} Lumi{\`e}re Lyon 2},
  \bibinfo{address}{France} (\bibinfo{year}{2015}).

\bibitem[{\citenamefont{Rose}(2014)}]{firstdbm}
\bibinfo{author}{\bibfnamefont{G.}~\bibnamefont{Rose}},
  \emph{\bibinfo{title}{{First ever DBM trained using a quantum computer}}},
  \bibinfo{howpublished}{\url{https://dwave.wordpress.com/2014/01/06/first-ever-dbm-trained-using-a-quantum-computer/}}
  (\bibinfo{year}{2014}), \bibinfo{note}{[Online; accessed 22-October-2015]}.

\bibitem[{\citenamefont{Raymond et~al.}(2016)\citenamefont{Raymond, Yarkoni,
  and Andriyash}}]{Raymond-DWave-2016}
\bibinfo{author}{\bibfnamefont{J.}~\bibnamefont{Raymond}},
  \bibinfo{author}{\bibfnamefont{S.}~\bibnamefont{Yarkoni}}, \bibnamefont{and}
  \bibinfo{author}{\bibfnamefont{E.}~\bibnamefont{Andriyash}},
  \bibinfo{journal}{arXiv:1606.00919}  (\bibinfo{year}{2016}).

\bibitem[{\citenamefont{Schneidman et~al.}(2006)\citenamefont{Schneidman,
  Berry, Segev, and Bialek}}]{Schneidman-Nature-2006}
\bibinfo{author}{\bibfnamefont{E.}~\bibnamefont{Schneidman}},
  \bibinfo{author}{\bibfnamefont{M.~J.} \bibnamefont{Berry}},
  \bibinfo{author}{\bibfnamefont{R.}~\bibnamefont{Segev}}, \bibnamefont{and}
  \bibinfo{author}{\bibfnamefont{W.}~\bibnamefont{Bialek}},
  \bibinfo{journal}{Nature} \textbf{\bibinfo{volume}{440}},
  \bibinfo{pages}{1007} (\bibinfo{year}{2006}).

\bibitem[{\citenamefont{M\'ezard and Mora}(2009)}]{Mezard-Mora-2009}
\bibinfo{author}{\bibfnamefont{M.}~\bibnamefont{M\'ezard}} \bibnamefont{and}
  \bibinfo{author}{\bibfnamefont{T.}~\bibnamefont{Mora}},
  \bibinfo{journal}{Journal of Physiology-Paris}
  \textbf{\bibinfo{volume}{103}}, \bibinfo{pages}{107 } (\bibinfo{year}{2009}),
  ISSN \bibinfo{issn}{0928-4257}, \bibinfo{note}{{Neuromathematics of Vision}}.

\bibitem[{\citenamefont{Ricci-Tersenghi}(2012)}]{Ricci-Tersenghi-JSTAT-2012}
\bibinfo{author}{\bibfnamefont{F.}~\bibnamefont{Ricci-Tersenghi}},
  \bibinfo{journal}{Journal of Statistical Mechanics: Theory and Experiment}
  \textbf{\bibinfo{volume}{2012}}, \bibinfo{pages}{P08015}
  (\bibinfo{year}{2012}).

\bibitem[{\citenamefont{Nguyen and Berg}(2012)}]{Chau-JSTAT-2012}
\bibinfo{author}{\bibfnamefont{H.~C.} \bibnamefont{Nguyen}} \bibnamefont{and}
  \bibinfo{author}{\bibfnamefont{J.}~\bibnamefont{Berg}},
  \bibinfo{journal}{Journal of Statistical Mechanics: Theory and Experiment}
  \textbf{\bibinfo{volume}{2012}}, \bibinfo{pages}{P03004}
  (\bibinfo{year}{2012}).

\bibitem[{\citenamefont{Aurell and Ekeberg}(2012)}]{Aurell-PRL-2012}
\bibinfo{author}{\bibfnamefont{E.}~\bibnamefont{Aurell}} \bibnamefont{and}
  \bibinfo{author}{\bibfnamefont{M.}~\bibnamefont{Ekeberg}},
  \bibinfo{journal}{Phys. Rev. Lett.} \textbf{\bibinfo{volume}{108}},
  \bibinfo{pages}{090201} (\bibinfo{year}{2012}).

\bibitem[{\citenamefont{Decelle and
  Ricci-Tersenghi}(2015)}]{Decelle-arXiv-2015}
\bibinfo{author}{\bibfnamefont{A.}~\bibnamefont{Decelle}} \bibnamefont{and}
  \bibinfo{author}{\bibfnamefont{F.}~\bibnamefont{Ricci-Tersenghi}},
  \bibinfo{journal}{arXiv:1501.03034}  (\bibinfo{year}{2015}).

\end{thebibliography}
\end{document}